\documentclass[%
 reprint,
superscriptaddress,
 amsmath,amssymb,
 aps,
 pre,
floatfix,
]{revtex4-2}

\usepackage{graphicx}% Include figure files
\usepackage{dcolumn}% Align table columns on decimal point
\usepackage{bm}% bold math
\usepackage{nicefrac}% nice fractions
\usepackage[textsize=tiny]{todonotes}
\usepackage{hyperref}% add hypertext capabilities
\usepackage{upgreek}
\usepackage{textgreek}
\usepackage[english]{babel}
\usepackage{amsmath}
\usepackage{verbatim}

\newcommand{\tmop}[1]{\ensuremath{\operatorname{#1}}}
\renewcommand{\eqref}[1]{Eq.~(\ref{#1})}

\newcommand{\figref}[1]{Fig.~\ref{#1}}
\newcommand{\secref}[1]{Sec.~\ref{#1}}
\newcommand{\appref}[1]{Appendix~\ref{#1}}

\newcommand{\kT}{k_{\text{B}}T}

\begin{document}

\title{Predicting the morphology of multiphase biomolecular condensates from protein interaction networks}

\author{Tianhao Li}
\affiliation{
 Department of Chemistry, Princeton University, Princeton, New Jersey 08544, USA\\
}
\author{William M. Jacobs}
 \email{wjacobs@princeton.edu}
\affiliation{
 Department of Chemistry, Princeton University, Princeton, New Jersey 08544, USA\\
}

\begin{abstract}
Phase-separated biomolecular condensates containing proteins and RNAs can assemble into higher-order structures by forming thermodynamically stable interfaces between immiscible phases. Using a minimal model of a protein/RNA interaction network, we demonstrate how a ``shared'' protein species that partitions into both phases of a multiphase condensate can function as a tunable surfactant that modulates the interfacial properties. We use Monte Carlo simulations and free-energy calculations to identify conditions under which a low concentration of this shared species is sufficient to trigger a wetting transition. We also describe a numerical approach based on classical density functional theory to predict concentration profiles and surface tensions directly from the model protein/RNA interaction network. Finally, we show that the wetting phase diagrams that emerge from our calculations can be understood in terms of a simple model of selective adsorption to a fluctuating interface. Our work shows how a low-concentration protein species might function as a biological switch for regulating multiphase condensate morphologies.
\end{abstract}

\maketitle

\section{Introduction}

Intracellular biomolecular mixtures can spatially organize into complex, self-assembled compartments via phase separation~\cite{Hyman14,BOEYNAEMS2018420,wheeler2018controlling}.
Such structures are referred to as \textit{biomolecular condensates}, since they form by spontaneously condensing biomolecular components, such as proteins and RNAs, into liquid-like compartments that are not enclosed by a membrane~\cite{banani2017}.
In many instances, condensates have been observed to assemble further into higher-order \textit{multiphasic} structures, in which multiple immiscible condensates form stable shared interfaces~\cite{cajal1999,shav2005nuclear,jcb2005}. Common multiphase morphologies include ``core--shell'' architectures, in which one condensate is completely surrounded by a second condensate, and ``docked'' architectures, in which condensed droplets attach to the surface of another condensate~\cite{science.aaf4382}.
The morphologies of many multiphase condensates appear to be related to their biological functions, such as the sequential processing of rRNA transcripts during ribogenesis within core--shell nucleoli~\cite{boisvert2007multifunctional}, and the sharing of various biomolecular components between docked stress granule and P-body condensates that are involved in regulating mRNA metabolism and translation~\cite{jcb2005,decker2012p,YOUN2019}.
It is therefore important to understand how the morphologies of multiphase condensates are controlled at a molecular level.

The formation of biomolecular condensates is widely considered to be a consequence of near-equilibrium, thermodynamically driven phase separation~\cite{Cliff2009,brangwynne2011active,berry2018physical}.
Within this thermodynamic framework, a multicomponent system evolves to minimize its overall free energy by phase-separating and adjusting the contact areas between different phases.
The equilibrium morphology of a multiphase system is thus governed by the relationships among surface tensions between pairs of phases and the volume fractions of the phases~\cite{mao2019phase,mao2020designing}.
Recent theoretical studies have demonstrated that surface tensions in multicomponent mixtures, and consequently multiphase morphologies, can be controlled by changing either the effective pairwise interactions between the biomolecular components~\cite{mao2020designing} or the stoichiometry of multicomponent condensates that are stabilized by heterotypic interactions~\cite{pyo2022surface}.
However, changing condensate morphologies via these mechanisms entails substantial changes to the state of the system, since the molecular properties and/or concentrations of the components that comprise the \textit{bulk} of the phase-separated condensates must be altered.

By contrast, surface tensions can be tuned by making comparably small perturbations to molecular components that adsorb to condensate \textit{interfaces}~\cite{folkmann2021regulation,erkamp2023adsorption,sanchez2021size}.
In principle, tuning the concentrations and affinities of surfactant-like components can control multiphase condensate morphologies with minimal changes to the state of an intracellular mixture, analogous to methods used to engineer multiphase emulsions~\cite{sheth2020multiple}.
A key example is provided by a recent study~\cite{sanders2020competing} of stress granules (SGs) and P-bodies (PBs), which assemble into a docked multiphase architecture under stressed conditions in human cells~\cite{YOUN2019}.
Importantly, this study suggested that small changes to the concentrations of specific proteins---in particular, those with affinities for proteins in each of the coexisting SG and PB phases---can trigger a transition between docked and dispersed condensates~\cite{sanders2020competing}. 
Although the localization of these particular proteins to the SG/PB interface has not been confirmed experimentally, this example suggests that molecular components with affinities for the constituents of multiple distinct condensates can alter the morphologies of multiphase condensates.

\begin{figure*}[t]
  \includegraphics[width=\textwidth]{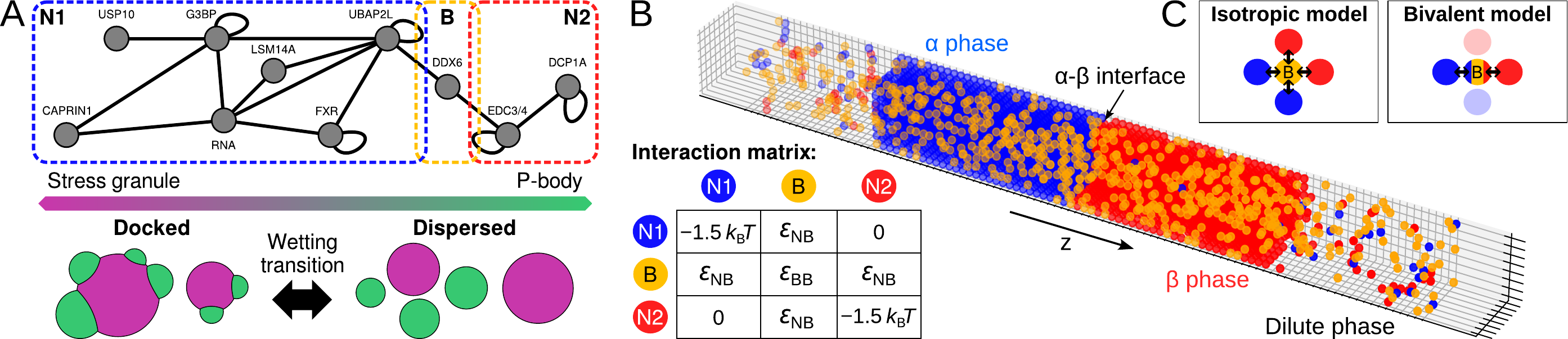}  \caption{\textbf{Minimal model of multiphase condensates.}
    (A)~Schematic of the protein/RNA interaction network that governs stress granule (SG)/P-body (PB) multiphase condensates~\cite{sanders2020competing}.  \textit{Top:} Nodes indicate protein/RNA species, while edges indicate either homotypic interactions (i.e., self associations) or heterotypic interactions (i.e., associations between different species).  We coarse-grain the network into N1, B, and N2 species.  The approximate partitioning of various species into the SG/PB condensates is suggested by the color bar below.  \textit{Bottom:} SG/PB condensates form either docked or dispersed morphologies, depending on the interactions specified by the network and the concentrations of the protein/RNA species.
    (B)~A typical configuration of the coarse-grained direct-coexistence simulations showing the $\alpha$ and $\beta$ condensed phases, composed primarily of the N1 (blue) and N2 (red) species, respectively, and a coexisting dilute phase.  The low-concentration B species (gold) is dispersed throughout the simulation box.  The interaction matrix, $\bm\epsilon$, for the nearest-neighbor intermolecular interactions is shown in the inset.
    (C)~Heterotypic interactions between B and node species either follow an isotropic model (left), in which all nearest-neighbor contacts contribute an interaction energy $\epsilon_{\text{NB}}$ (indicated by arrows), or a bivalent model (right), in which only oppositely positioned patches (spherical caps) on the B molecule interact with specific node species according to $\epsilon_{\text{NB}}$.
    \label{fig:1}}
\end{figure*}

In this article, we investigate this proposed mechanism for switching between morphologies of multiphase condensates.
We focus on the transition between non-wetting and partial wetting morphologies, which we refer to as the \textit{wetting transition} for brevity~\footnote{We note that in other usage, the wetting transition refers to the transition between partial wetting and complete wetting~\cite{deGennes_wetting}}. 
Specifically, we use a minimal model of a multicomponent mixture to derive design rules for controlling wetting transitions via low-concentration ``programmable surfactant'' proteins, which interact selectively with the constituents of two immiscible condensates.
We first introduce a simulation approach for computing the wetting transition between docked and dispersed morphologies.
We then develop a complementary theoretical approach based on classical density functional theory, which reproduces our simulation results semi-quantitatively.
Both approaches predict that relatively low concentrations of surfactant-like proteins can trigger a wetting transition between docked and dispersed morphologies under specific conditions.
Finally, we describe a qualitative theory that predicts the key features of this wetting transition and establishes rational design rules for understanding the behavior of programmable surfactants in multicomponent biomolecular mixtures.
Taken together, our results show how programmable surfactants can act as low-concentration molecular switches for regulating biological processes by controlling the morphologies of multiphase condensates.

\section{Minimal model of a programmable surfactant}

Our model is motivated by the multicomponent SG/PB system studied in Ref.~\cite{sanders2020competing}.
At a molecular level, the formation of the immiscible SG and PB condensates is dictated by the interactions among the constituent protein and mRNA components.
In this system, the relevant intermolecular interactions can be described by a protein/RNA interaction network (\figref{fig:1}A), in which nodes represent proteins, protein complexes, or RNA, and edges indicate attractive interactions between species.

To reduce the complexity, we coarse-grain the endogenous SG/PB protein/RNA interaction network to three explicit molecular components based on the organization of the network.
We refer to these coarse-grained species as ``Node 1'' (N1), ``Bridge'' (B), and ``Node 2'' (N2) throughout this work.
N1 and N2 are the majority components of the immiscible $\alpha$ and $\beta$ condensed phases, respectively, that form as a result of attractive homotypic interactions (\figref{fig:1}B).
The $\alpha$ and $\beta$ phases coexist with a dilute phase (D), which represents the cytosol in our implicit-solvent model.
For simplicity, we consider a three-dimensional lattice-gas model in which the N1, N2, and B species occupy individual lattice sites on a cubic lattice with lattice constant $\sigma$.
The homotypic and heterotypic interactions among these species are summarized in a pairwise interaction matrix $\bm\epsilon$~\cite{jacobs2023theory} (\figref{fig:1}B).
Details of the simulation approach are provided in \appref{app:simulations}.

Bridge molecules (B) represent a ``shared'' species that interacts with the majority components of the $\alpha$ and $\beta$ phases via attractive heterotypic interactions.
In this work, we consider two distinct models for the B species.
We first analyze an ``isotropic'' model in which B molecules interact with all nearest-neighbor N1 and N2 molecules; this model is most appropriate for describing highly multivalent protein and RNA species when the net interactions between pairs of molecules are weak compared to the thermal energy and can thus be approximated via isotropic pair interactions~\cite{jacobs2023theory}.
We then study an anisotropic ``bivalent'' model, in which B molecules interact with node molecules via specific patches (\figref{fig:1}C; see \secref{sec:bivalent}).
In this case, each of the two patches located on opposite sides of a B molecule only engages in heterotypic interactions with a specific node species.
We show that these two models yield qualitatively similar results for the wetting transition, suggesting that our minimal model serves as a reasonable approximation for many systems with anisotropic and multivalent interactions.

\section{Computing wetting transitions via molecular simulation}
\label{sec:simulation}

In this section, we introduce a Monte Carlo technique for computing the wetting transition between docked and dispersed morphologies of a multiphase condensate.
Using direct-coexistence Monte Carlo simulations in the slab geometry~\cite{panagiotopoulos2000monte} (\figref{fig:1}B), we determine the potential of mean force (PMF) between a pair of condensed phases in the canonical ensemble.
We then show how the properties of these PMFs can be related to the equilibrium morphology of a macroscopic multiphase system in the thermodynamic limit.

\subsection{Potential of mean force (PMF) calculations}

\begin{figure}[t]
  \includegraphics[width=\columnwidth]{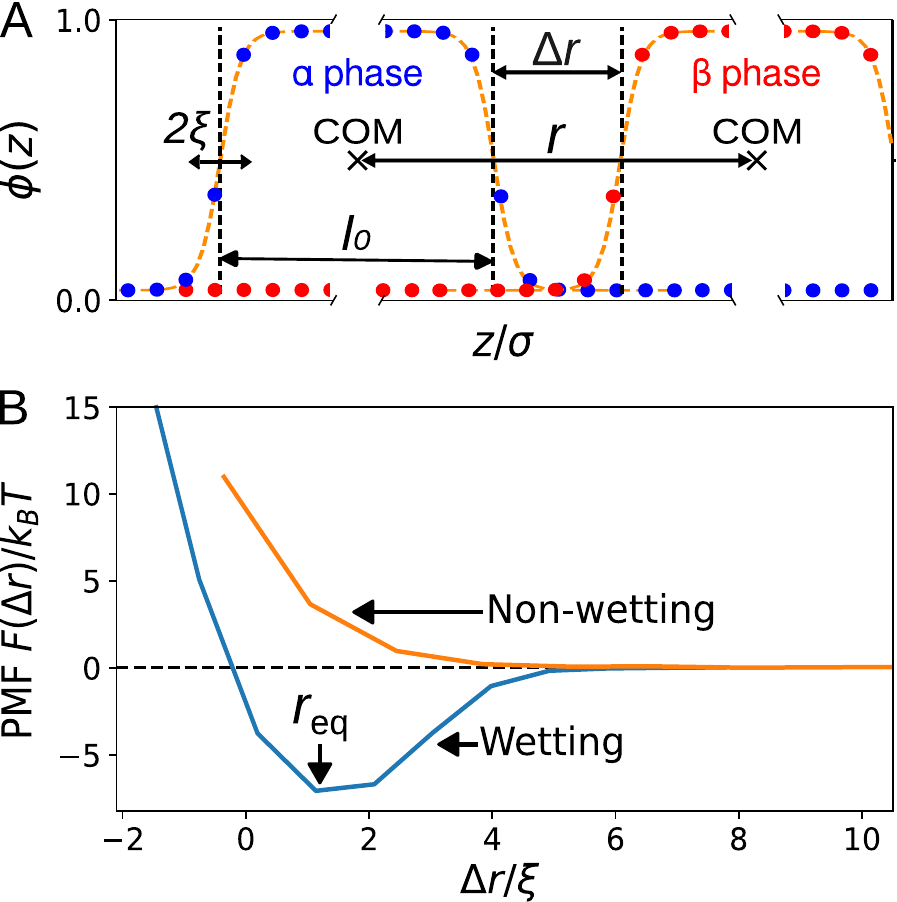}
  \caption{\textbf{Characterizing interfaces using Monte Carlo simulations and umbrella sampling.}
    (A)~Definitions of the characteristic interfacial width parameter, $\xi$, the width of a single condensed phase, $l_0$, and the distances between the phase centers of mass (COMs), $r$, and the Gibbs dividing surfaces (vertical dashed lines), $\Delta r$.
    (B)~The potential of mean force (PMF) as a function of the dimensionless distance parameter $\Delta r/ \xi$.  Example PMFs are shown for typical wetting (orange) and non-wetting (blue) scenarios.  Statistical errors are smaller than the line width.
    \label{fig:2}}
\end{figure}

To efficiently sample both wetting and non-wetting configurations, we perform umbrella sampling~\cite{TORRIE1977} by applying a harmonic biasing potential to the center-of-mass (COM) distance between the $\alpha$ and $\beta$ condensates (\figref{fig:2}A).
We first define the $\alpha$ and $\beta$-phase regions, $S_{\alpha/\beta}$, as the cross-sections along the $z$ axis of the simulation box with N1 or N2 volume fractions, $\phi_{\text{N1}/\text{N2}}(z)$, greater than $\phi^*$: $S_{\alpha/\beta} \equiv \{z \ \vert \ \phi_{\text{N1}/\text{N2}}(z)>\phi^*\}$.
We find that using a threshold of $\phi^* = 0.56$ reduces the effects of density fluctuations near the interfaces and thus improves the efficiency of our calculations; however, this choice has no significant effect on the results, as long as $\phi^*$ is situated between the molecular volume fractions of the bulk condensed and dilute phases.
We then compute the COM distance, $r$, based on the center of mass of the N1 or N2 molecules within the $\alpha$ or $\beta$ phases, respectively.
The COM distance is therefore $r \equiv \langle z \rangle^{(\beta)} - \langle z \rangle^{(\alpha)}$, where $\langle z \rangle^{(\alpha/\beta)} \equiv \int_{z \in S_{\alpha/\beta}} z\,dz / \int_{z \in S_{\alpha/\beta}} dz$, and we use the minimum-image convention to define distances given the periodic boundary conditions.

Our aim is to compute the PMF
\begin{equation}
    F(r) \equiv -\kT \ln \frac{p(r)}{p(r_\text{ref})},
\label{eq:pmf}
\end{equation}
where $p(r)$ is the probability of finding the $\alpha$ and $\beta$ condensed phases separated by a COM distance equal to $r$.
We choose a reference point for the PMF calculations where the interactions between the droplets are expected to be negligible (see \appref{app:simulations}).
Following the canonical umbrella sampling approach~\cite{TORRIE1977}, we apply a harmonic biasing potential, $(k_i/2)(r - r_{0,i})^2$, to constrain the COM distance near a target distance $r_{0,i}$.
Independent simulations, indexed by $i = 1, \ldots, M$, are used to sample near target COM distances at intervals of one~$\sigma$.
The spring constants $\{k_i\}$ are chosen to ensure that the probability distributions, $p_i(r)$, sampled from simulations at adjacent target distances overlap~\cite{frenkel2023understanding}.
In production simulations, we calculate the COM distance every 10 MC sweeps and record 200,000 samples for every target distance $\{r_{0,i}\}$. 
Finally, we utilize the multistate Bennett acceptance ratio (MBAR) method~\cite{Shirts_2008} to combine samples from the $M$ independent biased simulations and to obtain the unbiased PMF given by \eqref{eq:pmf}.
Example PMF calculations are shown in \figref{fig:2}B.

\subsection{Morphology predictions using the PMF}

The PMF defined via \eqref{eq:pmf} reflects the propensity for the $\alpha$ and $\beta$ condensed phases to assemble into a docked configuration, since the minimum value of the PMF corresponds to the equilibrium distance between the Gibbs dividing surfaces.
To assist in interpreting the PMF calculations, we characterize the distance between the $\alpha$ and $\beta$-phase interfaces by defining a dimensionless distance parameter $\Delta r/ \xi \equiv (r-l_0)/\xi$ (\figref{fig:2}A).
$\Delta r / \xi$ is equal to zero when the Gibbs dividing surfaces of the two condensed phases are in direct contact, whereas $\Delta r / \xi \gg 1$ indicates that the distance between the condensed-phase interfaces is large compared to the typical interfacial width.

Two representative PMFs for wetting and non-wetting scenarios are shown in \figref{fig:2}B.
In the non-wetting case, the PMF is non-negative, indicating a net repulsion between the $\alpha$ and $\beta$ phases.
We find that the PMF begins to increase as $\Delta r / \xi$ decreases below $\sim 4$, suggesting that the fluctuating interfaces interact well before the Gibbs dividing surfaces come into contact.
By contrast, the PMF has a clear minimum in the wetting case.
The negative values of the PMF at COM distances in the range $0 \lesssim \Delta r / \xi \lesssim 4$ indicate a net attraction between the $\alpha$ and $\beta$ condensed phases that also occurs over a length scale comparable to that of the interfacial fluctuations.

The PMFs from finite-size simulations enable us to predict the multiphase morphology of the systems in the thermodynamic limit (see \appref{app:scaling}).
The PMF minimum for a wetting interface (\figref{fig:2}B) is proportional to the cross-sectional area, $A$, where the constant of proportionality is the difference between the surface tensions at the $\alpha$--$\beta$ interface, $\gamma_{\alpha\beta}$, and the dilute--condensed interfaces, $\gamma_{\alpha\text{D}}$,
\begin{equation}
    F(r_{\text{eq}}) = -A (-\gamma_{\alpha \beta} + 2 \gamma_{\alpha \text{D}}).
    \label{eq:PMFwell}
\end{equation}
\eqref{eq:PMFwell} is supported by the finding that the PMF profiles scale linearly with $A$ in our simulations in both wetting and non-wetting cases (see \figref{fig:scaling}).
Thus, the multiphase morphology is determined solely by the PMF, or equivalently by the surface-tension difference $-\gamma_{\alpha \beta} + 2 \gamma_{\alpha \text{D}}$, in the thermodynamic limit.

\section{Predicting multiphase morphology with classical density functional theory}
\label{sec:cdft}

In this section, we develop a complementary approach for predicting multiphase condensate morphologies using the framework of classical density functional theory (CDFT).
We first show how to compute equilibrium concentration profiles for the isotropic bridge model.
We then discuss how these calculations can be used to predict the wetting transition between docked and dispersed morphologies of a multiphase condensate.

\subsection{Classical density function theory (CDFT)}

Assuming a regular solution model~\cite{Hildebrand1929}, the Helmholtz free-energy density, $f_0$, of the multicomponent lattice gas can be written as
\begin{equation}
  \frac{f_0 \sigma^3}{\kT} = \sum_i \phi_i \ln \phi_i + \frac{1}{2} \sum_{i,j} \phi_i \chi_{ij} \phi_j,
  \label{eq:free-energy}
\end{equation}
where the sums run over all $N$ molecular components as well as the implicit solvent.
The molecular volume fractions are constrained by $\sum^{N}_{i = 0} \phi_i = 1$, where $\phi_{0}$ represents the volume fraction of the solvent.
The interaction matrix $\chi_{ij}$ is related to the nearest-neighbor interaction energy, $\epsilon_{ij}$, by $\chi_{ij} \equiv (z/2\kT) (2\epsilon_{ij}-\epsilon_{ii}-\epsilon_{jj})$, where the lattice coordination number is $z=6$.
For the interaction matrix shown in \figref{fig:1}B, \eqref{eq:free-energy} predicts three coexisting phases, $\alpha$, $\beta$, and D, when the B-species volume fraction, $\phi_{\text{B}}$, is small, in agreement with our Monte Carlo simulation results.
Details of the phase-coexistence calculation and its numerical implementation are provided in \appref{app:coexistence}.
This mean-field regular solution model provides an adequate description of the bulk phases under these conditions, which are chosen to be sufficiently far from the critical points of the $\alpha$ and $\beta$ phases.

In the grand canonical ensemble, we express the grand-potential functional in terms of the square-gradient approximation~\cite{HANSEN2013203},
\begin{equation}
  \Omega [\vec{\phi}(z)] = A \!\! \int \! \left[ \omega_0 [\vec{\phi}(z)] +
  \frac{1}{2} \sum_{i, j} \phi'_i (z) m_{i j} \phi'_j (z) \right] \! d z,
  \label{eq:grandpotential}
\end{equation}
assuming planar interfaces as in our simulations.
This approximation assumes that inhomogeneities, such as interfaces between coexisting phases, vary slowly in space (i.e., over long wavelengths) due to the absence of higher-order derivatives~\cite{HANSEN2013203}.
$\Omega$ is a functional of the molecular volume fractions, $\vec{\phi}(z)\equiv(\phi_{\rm N1}(z),\phi_{\rm B}(z),\phi_{\rm N2}(z))^\top\!$, in a system at fixed chemical potentials, $\vec{\mu}\equiv(\mu_{\rm N1},\mu_{\rm B},\mu_{\rm N2})^\top\!$.
For interfacial property calculations, the chemical potentials are determined from the aforementioned coexistence conditions, such that the bulk phases far from an interface are in coexistence.
The first term in the integrand of \eqref{eq:grandpotential} is the local grand-potential density, $\omega_0 \equiv f_0 - \sum_i \mu_i \phi_i$.
The second term, involving derivatives of volume fractions with respect to $z$, $\phi'_i (z)$, represents the excess grand potential due to an inhomogeneity.

We approximate the coefficients of the square-gradient term, $\bm{m} = \{ m_{i j} \}$, by again assuming that the inhomogeneity is small in amplitude and varies slowly in space.
In this case, the $m_{i j}$ coefficients are determined from second derivatives of the free-energy density~\cite{cahn1958free,HANSEN2013203}.
For the free-energy density given in \eqref{eq:free-energy}, these conditions imply that
\begin{equation}
  m_{i j} = - \sigma^{-1} \epsilon_{i j}
  \label{eq:mmatrix}
\end{equation}
is a constant, concentration-independent matrix~\cite{HANSEN2013203,cahn1958free}.
However, in our multicomponent model, this matrix may not be positive-semidefinite, as required by the long-wavelength assumption underlying the square-gradient approximation.
If the coefficient matrix instead has negative eigenvalues, then large-amplitude inhomogeneities act to decrease the grand potential, leading to unphysical negative surface tensions and numerical instabilities.
Qualitatively, this scenario tends to occur when heterotypic interactions out-compete one or more homotypic interactions.
We propose that the square-gradient approximation can nonetheless be applied to multicomponent solutions in such scenarios by regularizing the $\bm{m}$-matrix.
We therefore perform an eigenvalue decomposition of \eqref{eq:mmatrix}, replace the negative eigenvalues (if there are any) with zeroes, and reconstruct the regularized low-rank~\cite{eckart1936rank} $\bm{m}$-matrix for use in \eqref{eq:grandpotential}.
After regularization, the fluctuation modes represented by the eigenvectors with zero eigenvalues do not contribute to the square gradient term in \eqref{eq:grandpotential}. 

The equilibrium interfaces between coexisting phases are determined by minimizing the grand-potential functional,
\begin{equation}
  \frac{\delta \Omega [\vec{\phi}(z)]}{\delta \phi_i} = 0,
\label{eq:euler-lagrange}
\end{equation}
which yields the equilibrium molecular volume-fraction profiles, $\vec\phi_{\text{eq}}(z)$.
Despite the shortcomings of the long-wavelength assumption, we find that our approach for regularizing the $\bm{m}$-matrix leads to semi-quantitative predictions for the molecular volume-fraction profiles across a wide variety of conditions, as we show in \secref{sec:enrichment}.

\subsection{Morphology predictions based on CDFT}

To predict the equilibrium morphology of multiphase condensates, we calculate the excess free-energy profile, $\Delta \omega (z)$, in the vicinity of an interface between bulk phases,
\begin{equation}
  \Delta \omega (z) = \omega_0 [\vec\phi_{\text{eq}}(z)] - \omega_0^{(\text{D})} + \frac{1}{2}
  \sum_{i, j} \phi'_{\text{eq},i} (z) m_{i j} \phi'_{\text{eq},j} (z),
  \label{eq:excess-free-energy}
\end{equation}
where $\omega^{(\text{D})}_0$ is the grand potential of the bulk dilute phase.
The associated surface tension,
\begin{equation}
  \gamma = \int \Delta \omega (z) d z,
  \label{eq:surface-tension}
\end{equation}
is then obtained by integrating the excess free-energy profile across the interface.
Finally, we compute the surface-tension difference $-\gamma_{\alpha \beta} + 2 \gamma_{\alpha \text{D}}$ by applying \eqref{eq:surface-tension} to both the $\alpha$--$\beta$ and the $\alpha$--D interfaces.

We emphasize that the Euler--Lagrange equation specified by \eqref{eq:euler-lagrange} must be solved numerically for our multicomponent model, since the concentration of the B species can vary non-monotonically across an interface.
In practice, this can be achieved by minimizing $\Omega[\vec{\phi}(z)]$ via gradient descent.
Details regarding our implementation of this numerical scheme, as well as criteria for assessing convergence, are presented in \appref{app:euler-lagrange}.
As we show in \secref{sec:wetting}, this numerical approach predicts wetting transitions in qualitative agreement with our Monte Carlo simulation results.
By contrast, assuming that the molecular volume-fraction profiles follow linear paths through concentration space~\cite{mao2019phase,mao2020designing,sanders2020competing} predicts non-wetting behavior for a wide range of conditions, which is at odds with our simulation results.
We discuss this approximation, as well as the relationship between our method and an alternative ``minimum free-energy path'' (MFEP) approximation, in \appref{app:CH}.

\begin{figure*}[t]
  \includegraphics[width=\textwidth]{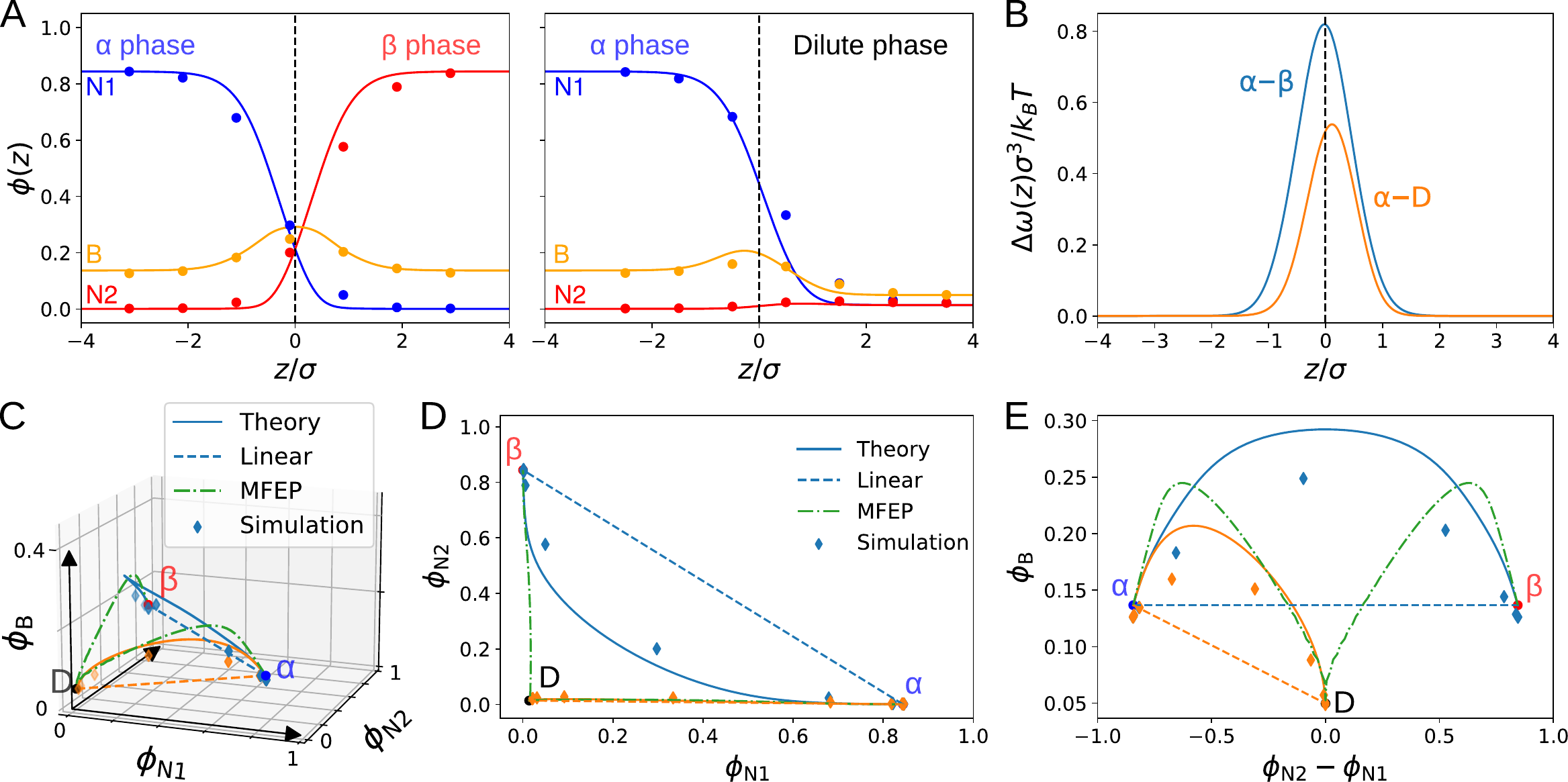}
  \caption{\textbf{Enrichment of the programmable surfactant at condensate interfaces.}
    Here, calculations are performed using the isotropic bridge model.
    (A)~CDFT predictions (solid lines) closely agree with simulation measurements (points) of the equilibrium concentration profiles at $\alpha$--$\beta$ (left) and $\alpha$--D (right) interfaces.  Vertical dashed lines indicate the Gibbs dividing surfaces.  Results are shown for a typical wetting case ($\epsilon_{\text{NB}} = -1\,\kT$, $\epsilon_{\text{BB}} = 0$, $\phi_\text{B} = 0.1$).
    (B)~Excess free-energy profiles corresponding to the CDFT profiles shown in \textbf{A}.
    (C--E)~Concentration profiles across the interfaces trace out paths in the three-dimensional (N1, N2, B) concentration space.  Comparisons are shown between the simulation results (points), the full CDFT theory (solid lines), the linear-path approximation (dashed lines), and the minimum-free-energy-path (MFEP) approximation (dash-dotted lines) for the cases shown in \textbf{A}.
  Simulation statistical errors are comparable to the symbol size.}
  \label{fig:CDFT}
\end{figure*}

\section{Controlling wetting transitions using a programmable surfactant}
\label{sec:results}

We now investigate how the ``programmable surfactant'' (B) species, which is shared between the $\alpha$ and $\beta$ condensates in \figref{fig:1}, controls the multiphase condensate morphology.
To this end, we first study the behavior of the isotropic model at different B-species volume fractions, $\phi_\text{B}$; heterotypic N--B binding affinities, $\epsilon_{\text{NB}}$; and homotypic B--B interaction strengths, $\epsilon_{\text{BB}}$.
We focus specifically on the low-$\phi_{\text{B}}$, weak-$\epsilon_{\text{NB}}$ regime, in which the compositions of the bulk $\alpha$ and $\beta$ phases are negligibly affected by the presence of the B species, as we expect that this regime is most relevant to the regulation of multiphase condensate morphologies in a biological context.
We then demonstrate that the bivalent model results in qualitatively similar behavior.

\subsection{Surfactant enrichment at wetting interfaces}
\label{sec:enrichment}

We first examine the correspondence between the interfacial concentration profiles predicted by simulations (see \secref{sec:simulation}) and CDFT (see \secref{sec:cdft}) under wetting and non-wetting conditions using the isotropic model (\figref{fig:CDFT}A).
In the wetting case, we estimate the equilibrium concentration profile at the $\alpha$--$\beta$ interface from simulations conducted with the biasing potential centered at the equilibrium COM distance, $r_{\text{eq}}$ (see, e.g., \figref{fig:2}B and \appref{app:simulations}).
In the non-wetting case, we examine the $\alpha$--D interface in the absence of the $\beta$ phase.
We define the Gibbs dividing surfaces by symmetry in the case of the $\alpha$--$\beta$ interface, and on the basis of $\phi_{\text{N1}}(z)$ in the case of the $\alpha$--D interface~\cite{HANSEN2013203}.
We generically find a slight but statistically significant enrichment of the B species at both $\alpha$--$\beta$ and $\alpha$--D interfaces when $\epsilon_{\text{NB}} < 0$.
This effect is greater at $\alpha$--$\beta$ interfaces under wetting conditions, as might be expected for a surfactant-like species that is attracted to both condensed phases.
Nonetheless, we note that even under wetting conditions, only a small fraction of all B molecules are located at the $\alpha$--$\beta$ interface.

\begin{figure}[t]
  \includegraphics[width=\columnwidth]{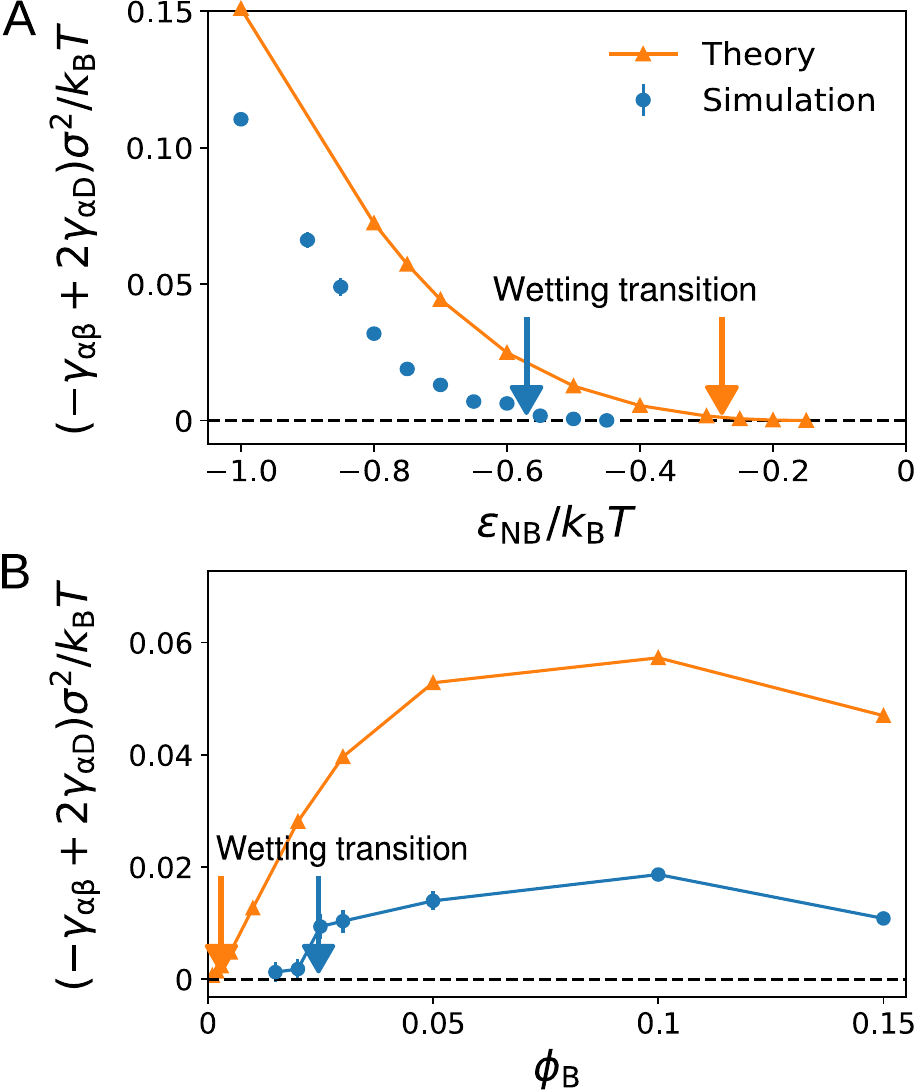}
  \caption{\textbf{Identifying surfactant-dependent wetting transitions.}
    (A)~The difference between the non-wetting and wetting surface tensions, $-\gamma_{\alpha\beta}+2\gamma_{\alpha\text{D}}$, as a function of $\epsilon_{\text{NB}}$ at constant $\phi_\text{B}=0.1$ using the isotropic bridge model with $\epsilon_{\text{BB}}=0$.  Results are shown for simulations (blue circles) and CDFT calculations (orange triangles).  Arrows indicate the locations of the wetting transition, where the surface-tension difference equals zero to within twice the statistical uncertainty, inferred from both methods.
    (B)~The surface-tension difference as a function of $\phi_{\text{B}}$ at constant $\epsilon_{\text{NB}} = -0.75\,\kT$.  Statistical uncertainties are comparable to the symbol size.
    }
  \label{fig:transition}
\end{figure}

Despite the approximations inherent to our CDFT approach, we find semi-quantitative agreement between the predicted and simulated concentration profiles (\figref{fig:CDFT}A).
The interfaces predicted by CDFT tend to exaggerate the B-species enrichment at the interface relative to the simulation results.
Furthermore, from CDFT calculations, we directly obtain predictions for the excess free energy across each interface, $\Delta \omega (z)$ (\figref{fig:CDFT}B).
In the case of the $\alpha$--D interface, the excess free energy profile is asymmetric about the Gibbs dividing surface, with the maximum shifted toward the dilute phase.

The behavior of the B species near each interface is more clearly seen in the three-dimensional (N1, N2, B) concentration space (\figref{fig:CDFT}C--E).
The enrichment of the B species relative to its concentration in either the condensed or dilute phase results in a marked deviation from the linear-path approximation (dashed lines in \figref{fig:CDFT}C--E).
Consequently, the excess free energy predicted by the full CDFT approach for the $\alpha$--$\beta$ interface shown in \figref{fig:CDFT}A--B is substantially lower than that predicted by this linear-path constraint.
Because the deviation from the linear-path approximation is greater for the $\alpha$--$\beta$ interface than the $\alpha$--D interface, the linear-path approximation tends to mischaracterize wetting conditions as non-wetting.
By contrast, MFEP calculations (dot-dashed lines in \figref{fig:CDFT}C--E) exaggerate the enrichment of B molecules at all interfaces, and we find that the MFEP between the $\alpha$ and $\beta$ phases in concentration space can actually pass through the dilute phase.
The full CDFT approach, which most closely matches the simulation results, is intermediate between these limiting cases, exhibiting a reduction of the N1 and N2 concentrations at the interface without passing through the dilute phase (\figref{fig:CDFT}D).
We stress that these predictions are dependent on our regularization approach for the $\bm{m}$-matrix (see \secref{sec:cdft}), without which CDFT would yield diverging interfacial fluctuations for the parameters used in \figref{fig:CDFT}.
Overall, these comparisons demonstrate the semi-quantitative accuracy of our CDFT approach and highlight shortcomings of the linear-path approximation (see \appref{app:CH}) in multicomponent settings.

\subsection{Computing the wetting transition}
\label{sec:wetting}

To determine how the equilibrium multiphase morphology changes with the concentration and heterotypic interactions of the B species, we compute the surface-tension difference $-\gamma_{\alpha \beta} + 2 \gamma_{\alpha \text{D}}$ using both simulation results and CDFT calculations (\figref{fig:transition}A--B). 
A positive surface-tension difference, $-\gamma_{\alpha \beta} + 2 \gamma_{\alpha \text{D}} > 0$, indicates a stable wetting interface between the $\alpha$ and $\beta$ phases.
From our simulation results, we compute this quantity based on the PMF minimum (\secref{sec:simulation}), and we identify the wetting transition where the PMF minimum becomes statistically indistinguishable from zero (blue arrows in \figref{fig:transition}A--B).
In our CDFT calculations, the $\alpha$--$\beta$ interface spontaneously relaxes to two dense--dilute interfaces when a non-wetting configuration is predicted, in which case we obtain a near-zero value for the surface-tension difference due to finite numerical precision (see \appref{app:euler-lagrange}).
We therefore identify the CDFT wetting transition by comparing the surface-tension difference to the numerical precision (orange arrows in \figref{fig:transition}A--B).

Our simulations and CDFT calculations predict qualitatively similar behavior for the surface-tension difference and the location of the wetting transition.
In particular, both simulation and theory predict that the surface-tension difference tends to increase, leading to a stable wetting configuration, with decreasing $\epsilon_\text{NB}$ and increasing $\phi_{\text{B}}$.
However, the wetting transition occurs at weaker N--B interactions and lower B concentrations in the CDFT theory.
This quantitative discrepancy likely arises from the mean-field and long-wavelength assumptions invoked in the CDFT theory.
Nonetheless, it is interesting that the equilibrium concentration profiles predicted by CDFT appear to be in much closer agreement with the simulation results (\figref{fig:CDFT}A) than the surface-tension differences (\figref{fig:transition}).
This comparison suggests that the key shortcoming of the CDFT theory lies in the neglect of interfacial fluctuations, which we expect to have stronger effects on surface free energies than average concentration profiles.

\subsection{Design rules for regulating multiphase condensate morphology via programmable surfactants}

\begin{figure}[t!]
  \includegraphics[width=\columnwidth]{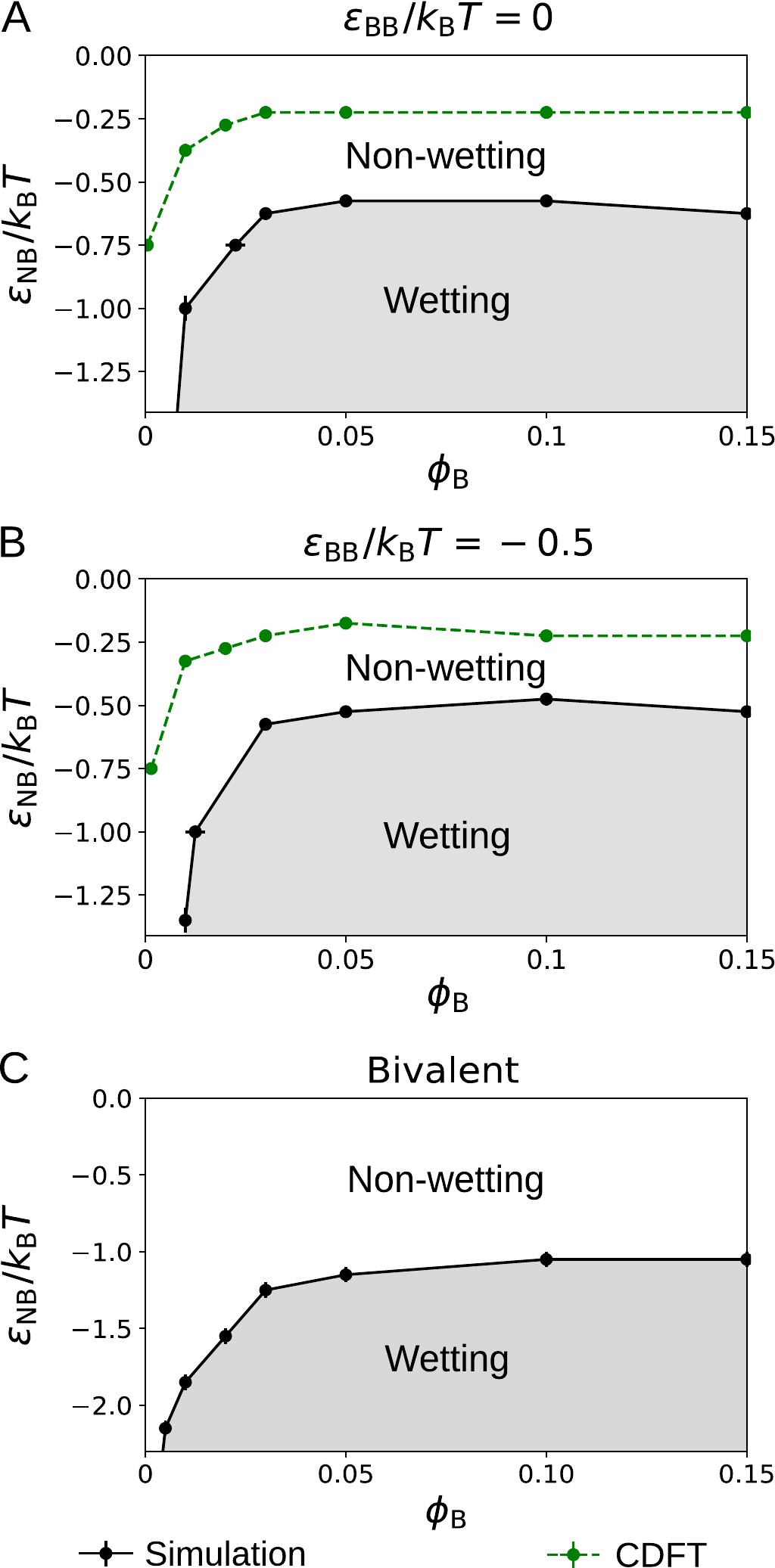}
  \caption{\textbf{Wetting phase diagrams for programmable surfactacts.}
    Phase diagrams are shown as a function of the B-species volume fraction, $\phi_{\text{B}}$, and the heterotypic N--B interaction energy, $\epsilon_{\text{NB}}$, for isotropic bridge models (A)~without homotypic interactions and (B)~with weak homotypic interactions ($\epsilon_{\text{BB}} = -0.5\,\kT$), and (C)~for the bivalent bridge model. Shaded regions indicate wetting conditions.  Phase boundaries are predicted from simulation results (black points) and CDFT surface-tension calculations (green points).}
  \label{fig:PD}
\end{figure}

The equilibrium multiphase morphology of the isotropic bridge model can be summarized in a wetting phase diagram (\figref{fig:PD}A).
Here we plot separate curves predicting the wetting transition based on simulations and the CDFT surface-tension difference in the absence of homotypic bridge interactions ($\epsilon_{\text{BB}}=0$).
The shaded region of parameter space below each curve in the $\epsilon_{\text{NB}}$--$\phi_{\text{B}}$ plane corresponds to an equilibrium wetting configuration, where a docked multiphase morphology is thermodynamically stable.
We also report the phase diagram for a model in which B molecules interact via weak homotypic interactions, such that $\epsilon_{\text{BB}} = -0.5\,\kT$ (\figref{fig:PD}B).
Although the CDFT approach underestimates the N--B interaction strength required to trigger the wetting transition, it captures the qualitative shape of the phase diagram both with and without homotypic interactions.

The wetting phase diagrams presented in \figref{fig:PD}A--B exhibit a number of striking features.
First, there is a minimum heterotypic interaction strength, $|\epsilon_{\text{NB}}^*|$, required for a stable wetting configuration.
Stable docked morphologies therefore cannot occur for $\epsilon_{\text{NB}} > \epsilon_{\text{NB}}^*$, regardless of the B-species concentration.
In this model, we find that $\epsilon_{\text{NB}}^* \approx -0.6\,\kT$, which is considerably weaker than the critical interaction strength of the cubic lattice gas model~\cite{jacobs2014phase}.
Second, the wetting transition passes through $\epsilon_{\text{NB}}^*$ at a finite B-species volume fraction.
This observation implies that the wetting transition is re-entrant for values of $\epsilon_{\text{NB}}$ close to $\epsilon_{\text{NB}}^*$, where increasing $\phi_{\text{B}}$ at a constant heterotypic interaction strength leads the system to transition from the non-wetting regime to the wetting regime, and then back to the non-wetting regime at high B-species concentrations.
Third, the phase boundary extends to low B-species volume fractions, on the order of only a few percent.
Importantly, at dilute B-species concentrations, the heterotypic interaction strength required to trigger the wetting transition weakens rapidly with increasing $\phi_{\text{B}}$.
By contrast, the phase boundary is comparably insensitive to $\phi_{\text{B}}$ near $\epsilon_{\text{NB}}^*$.

Finally, we observe that homotypic B--B interactions have only a minor quantitative effect on the wetting phase diagram.
This finding indicates that weak homotypic interactions among surfactant-like species play a secondary role in modulating multiphase condensate morphologies.
However, there are slight differences between the phase diagrams.
On one hand, introducing homotypic B--B interactions reduces the minimum required interaction strength, $|\epsilon_{\text{NB}}^*|$, by a small amount.
On the other hand, at dilute B-species concentrations, the wetting phase boundary is shifted to slightly stronger heterotypic N--B interactions.

Taken together, these observations establish general design rules for programmable surfactants.
Most importantly, our results indicate that relatively low concentrations of a surfactant-like species ($\phi_{\text{B}} \gtrsim 0.03$) and relatively weak heterotypic interactions ($\epsilon_{\text{NB}} \lesssim -0.6\,\kT$) are sufficient to trigger a wetting transition in a multicomponent, multiphase mixture.
We note that these phase diagrams are insensitive to changes in the concentrations of the N1 and N2 species, as these changes do not substantially affect the compositions of the bulk $\alpha$ and $\beta$ phases when the B species is dilute.
However, the equilibrium multiphase morphology may transition from partial wetting (i.e., a docked configuration) to complete wetting (i.e., a core--shell structure) when the $\alpha$ and $\beta$-phase volume fractions differ substantially~\cite{mao2019phase,mao2020designing}.

\subsection{Generalization to the bivalent bridge model}
\label{sec:bivalent}

We next investigate the behavior of the bivalent bridge model.
In this model, each anisotropic B molecule has two binding sites on opposite sides (\figref{fig:1}C).
One site selectively binds to N1 while the other binds to N2, each with interaction strength $\epsilon_{\text{NB}}$.
Thus, to establish a node--bridge interaction, a B molecule must be adjacent to a node molecule with the correct binding site pointing towards it, resulting in a larger entropic penalty for heterotypic interactions than in the isotropic model.

Despite these differences in the N--B binding rules, we find that the wetting phase boundary for the bivalent model is qualitatively similar to its isotropic-model counterparts (\figref{fig:PD}C).
In fact, we find that the enrichment of B molecules at $\alpha$--$\beta$ wetting interfaces is more pronounced with the bivalent model, since the B molecules are less miscible in the bulk condensed phases.
Minor differences arise since stronger N--B interactions are required to stabilize a wetting interface in the bivalent model due to the greater entropic penalty for heterotypic interactions.
As a result, the wetting transition is shifted lower in \figref{fig:PD}C.
We also find no evidence for a re-entrant wetting transition at bridge concentrations up to $\phi_\text{B}=0.3$.
Yet overall, the strong dependence of the wetting transition on the B-species volume fraction at low $\phi_\text{B}$ is preserved in the bivalent model.
This observation suggests that the switch-like mechanism for triggering a morphology change at low $\phi_{\text{B}}$ is a general feature of programmable surfactants, and is relatively insensitive to the details of the molecular model.

\subsection{Understanding programmable surfactant design rules using an adsorption model}
\label{sec:adsorption}

To gain a deeper understanding of these empirical design rules, we introduce a simple adsorption model that recapitulates the key features of the wetting phase diagrams in \figref{fig:PD}.
We examine the interplay between the parameters $\epsilon_{\text{NB}}$, $\epsilon_{\text{BB}}$, and $\phi_{\text{B}}$ by considering a Langmuir-like model~\cite{langmuir1918adsorption} in which the B species acts as the adsorbate.
We therefore assume that each fluctuating interface between a pair of phases can be described by a two-dimensional lattice gas with B-species occupancy $\phi_\text{B}^{(\text{i})}$.

\begin{figure}[t]
  \includegraphics[width=\columnwidth]{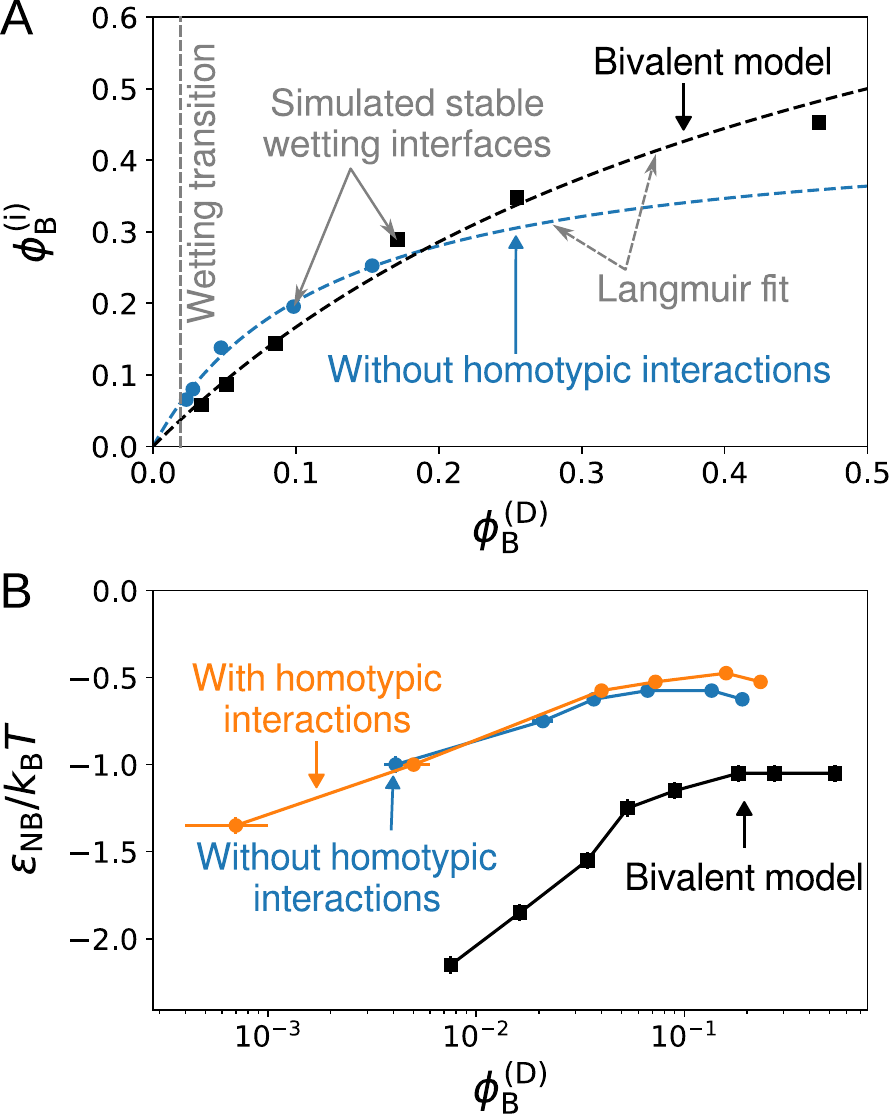}
  \caption{\textbf{An adsorption model predicts qualitative features of the wetting transition.}
    (A)~The enrichment of B molecules at the $\alpha$--$\beta$ interface can be described by a Langmuir isotherm.  Simulation measurements of the B-species enrichment for the isotropic model (blue points; $\epsilon_\text{NB}=-0.75k_\text{B}T$, $\epsilon_\text{BB}=0$) are fit to a Langmuir isotherm, \eqref{eq:iso}, with fitting parameter $a_{\alpha\beta}=2.1$ and an empirical scaling factor of $0.45$.  Measurements of the B-species enrichment in the bivalent model (black points; $\epsilon_\text{NB}=-1.6k_\text{B}T$) are fit to a Langmuir isotherm with fitting parameter $a_{\alpha\beta}=0.69$.
    (B)~Wetting phase boundaries as a function of the B-species volume fraction in the dilute phase. (Data are obtained from the same simulations as presented in \figref{fig:PD}, although the quantities being plotted, $\phi_{\text{B}}$ versus $\phi_{\text{B}}^{(\text{D})}$, are different.) The adsorption model predicts the nonmonotonicity of the isotropic-model phase boundary, the asymptotic behavior of all models at low $\phi_{\text{B}}^{(\text{D})}$, and the increase of the maximum N--B interaction on the phase boundary, $\epsilon_{\text{NB}}^*$, due to homotypic B--B interactions.}
  \label{fig:lm}
\end{figure}

We first consider the isotropic and bivalent models without homotypic B-species interactions ($\epsilon_\text{BB}=0$).
The surface excess grand potential, $\Omega^{\text{ex}}$, due to the presence of the interface~\cite{HANSEN2013203} takes the form
\begin{equation}
  \frac{\Omega^\text{ex}(\phi^{(\text{i})}_{\text{B}})\sigma^2}{A k_\text{B} T} = h(\phi^{(\text{i})}_{\text{B}}) -\frac{\Delta S(\phi^{(\text{i})}_{\text{B}})}{A k_\text{B}}- \frac{\mu_\text{B}}{k_\text{B} T} \phi^{(\text{i})}_{\text{B}},
  \label{eq:lm_excess}
\end{equation}
where $A$ is the interfacial area.
The first enthalpic term is linearly related to the occupied volume fraction in the mean-field approximation, $ h(\phi^{(\text{i})}_{\text{B}})=-a\phi^{(\text{i})}_{\text{B}}+b$.
For a surfactant-like adsorbate, $a$ is positive.
Meanwhile, $b$ represents the enthalpic penalty due to the creation of an interface from a bulk condensed phase and is independent of $\phi^{(\text{i})}_{\text{B}}$.
The entropic contribution, $\Delta S(\phi^{(\text{i})}_{\text{B}}) / A k_\text{B} = -\phi^{(\text{i})}_{\text{B}} \ln \phi^{(\text{i})}_{\text{B}} - (1-\phi^{(\text{i})}_{\text{B}}) \ln (1-\phi^{(\text{i})}_{\text{B}}) + s$, accounts for the in-plane configurational entropy of the adsorbed B molecules as well as the entropic penalty, $s$, due to capillary fluctuations.
Finally, $\mu_\text{B}$ is the B-species chemical potential.
Since the dense phase is in coexistence with the approximately ideal dilute phase, we have $\mu_\text{B}/k_\text{B} T \simeq \ln \phi^{(\text{D})}_\text{B}$, where $\phi^{(\text{D})}_\text{B}$ is the B-species volume fraction in the dilute phase.
By minimizing the surface excess grand potential, \eqref{eq:lm_excess}, with respect to $\phi^{(\text{i})}_{\text{B}}$, we arrive at a Langmuir adsorption isotherm for the B species,
\begin{equation}
  \phi^{(\text{i})}_\text{B,eq} = \frac{K}{1+K},
  \label{eq:iso}
\end{equation}
where $K \equiv \phi^{(\text{D})}_\text{B} e^a$.
This prediction agrees well with the enrichment of B molecules at $\alpha$--$\beta$ interfaces in our simulations (\figref{fig:lm}A).
In the case of the bivalent model, we obtain quantitative agreement by tuning the coefficient $a$.
For the isotropic model, we find it necessary to introduce an overall scaling factor when fitting to the Langmuir isotherm, \eqref{eq:iso}, since the diffuse interfaces are wider than $2\sigma$, making the monolayer assumption in the adsorption model less appropriate.

We now compute the surface tension in this model from the equilibrium surface excess grand potential,
\begin{equation}
  \gamma \equiv \frac{\Omega^\text{ex}(\phi^{(\text{i})}_\text{B,eq})}{A} = \frac{k_\text{B} T}{\sigma^2}[b - s - \ln(1+K)],
  \label{eq:gamma}
\end{equation}
and apply this formula  to both the $\alpha$--$\beta$ and the $\alpha$--D interfaces.
The $\phi^{\text{(i)}}_{\text{B}}$-independent enthalpic contribution must be the same regardless of the distance between the $\alpha$ and $\beta$-phase interfaces, so that ${b_{\alpha\beta} = 2b_{\rm \alpha D}}$.
By contrast, the entropic contribution due to capillary fluctuations depends on whether one or two distinct interfaces are present between the $\alpha$ and $\beta$ phases.
We therefore define the dimensionless entropic difference ${\Delta s \equiv  - s_{\alpha\beta}+2s_{\rm \alpha D}}$, which is necessarily positive and increases with the interfacial roughness.
In a wetting case, 
\begin{equation}  
  -\gamma_{\rm \alpha \beta}+2\gamma_{\rm \alpha D} = -\frac{k_\text{B} T}{\sigma^2}\left[\Delta s + \ln \frac{(1+K_{\rm \alpha D})^2}{1+K_{\rm \alpha \beta}}\right] > 0.
  \label{eq:difference}
\end{equation}
\eqref{eq:difference} predicts a wetting phase boundary that is quadratic with respect to $\phi^{(\text{i})}_{\text{B}}$ (see \appref{app:adsorption}), which indicates that wetting can only occur when~\cite{song2021review}
\begin{equation}
  a_{\rm \alpha D} - a_{\rm \alpha \beta}<\ln \frac{1-\sqrt{1-e^{-\Delta s}}}{2}.
  \label{eq:ltbs}
\end{equation}
Because the coefficients $a_{\rm \alpha D}$ and $a_{\rm \alpha \beta}$ reflect the enthalpic contribution due to the adsorption of B molecules, we assume that the left-hand side of \eqref{eq:ltbs} is roughly proportional to $\epsilon_{\text{NB}}$.
Moreover, by making the approximation ${a_{\rm \alpha \beta} \simeq 2 a_{\rm \alpha D}}$, which implies that a B molecule engages in twice as many N--B interactions at an $\alpha$--$\beta$ interface, we obtain a relation between $\phi^{(\text{D})}_{\text{B}}$ and $\Delta s$ at the minimum binding strength, $|\epsilon_{\text{NB}}^*|$, on the wetting phase boundary,
\begin{equation}
  \left.\phi^{(\text{D})}_\text{B}\right|^* =\frac{-(1-e^{-\Delta s})+\sqrt{1-e^{-\Delta s}}}{2}.
\label{eq:phi_ltbs_2}
\end{equation}
Importantly, the existence of a minimum binding strength $|\epsilon_{\text{NB}}^*|$, resulting from \eqref{eq:difference}, predicts a non-monotonic wetting phase boundary, with a re-entrant wetting transition at constant $\epsilon_{\text{NB}} < \epsilon_{\text{NB}}^*$, as observed in our isotropic-model simulations.
In the bivalent model, the entropic penalty to orient the B-molecule binding sites perpendicular to the $\alpha$--$\beta$ interface implies a substantially higher $\phi^{(\text{D})}_{\text{B}}$ for re-entrance, consistent with our simulations (see \appref{app:adsorption}).

Finally, in the low-concentration and high-affinity regime, $e^{a_{\alpha \beta}-a_{\alpha \text{D}}}\gg e^{\Delta s}$, we obtain an asymptotic formula for the low-concentration phase boundary,
\begin{equation}
  \ln \phi^{(\text{D})}_\text{B} + a_{\alpha\beta} = \ln(e^{\Delta s}-1).
  \label{eq:boundary}
\end{equation}
The logarithmic dependence on $\phi_{\text{B}}^{(\text{D})}$ in \eqref{eq:boundary} explains the sensitivity of the wetting transition to the B-species concentration under these conditions (\figref{fig:lm}B), where we see that $\partial a_{\alpha\beta} / \partial \ln \phi^{(\text{D})}_\text{B} \propto \partial \epsilon_{\text{NB}} (k_{\rm B} T)^{-1}/ \partial \ln \phi^{(\text{D})}_\text{B}$ is roughly constant in both the isotropic and bivalent bridge models.
The behavior in the low-concentration, high-affinity regime can therefore be interpreted as a competition between capillary fluctuations and the free-energy change when adsorbing a B molecule to the $\alpha$-$\beta$ interface.

We now consider the isotropic model with homotypic interactions ($\epsilon_{\text{BB}} < 0$).
In this scenario, the mean-field approximation for the enthalpic contribution in \eqref{eq:lm_excess} acquires an extra term $-a'(\phi_{\text{B}}^{(\text{i})})^2$, where $a' > 0$, that accounts for B--B interactions at the interface.
The chemical potential $\mu_{\text{B}}$ similarly picks up a term that is proportional to $\phi_{\text{B}}^{(\text{D})}$, since the B molecules can also attract one another in the dilute phase.
The effects of these modifications can then be predicted by perturbing the $\epsilon_{\text{BB}}=0$ results (see \appref{app:adsorption}).
In the low-concentration limit, both $\phi_{\text{B}}^{(\rm i)}$ and $\phi_{\text{B}}^{(\text{D})}$ are small, so that the asymptotic behavior given by \eqref{eq:boundary} remains unchanged; this prediction is confirmed by plotting the wetting phase boundary as a function of $\phi_{\text{B}}^{(\text{D})}$ in \figref{fig:lm}B.
However, near $\phi^{(\text{D})}_\text{B}|^*$, the perturbation due to $a'$ is non-negligible.
Specifically, turning on homotypic B--B interactions results in a change to the surface tension difference,
\begin{equation}
  \Delta (-\gamma_{\rm \alpha \beta}+2\gamma_{\rm \alpha D})|^*=\frac{2K^2_{\rm \alpha D}|^*}{(1+K^*_{\rm \alpha D})^2} a' > 0,
\label{eq:delta_gamma*}
\end{equation}
at the maximum N--B interaction, $\epsilon_{\text{NB}}^*$, on the phase boundary.
The sign of \eqref{eq:delta_gamma*} indicates that homotypic B--B interactions weaken the required N--B interaction strength, resulting in an increased $\epsilon_{\text{NB}}^*$.
This prediction also agrees with our simulation results (\figref{fig:lm}B).

In summary, this analytical model explains all the essential features of our wetting phase diagrams, including the re-entrant wetting transition and the low-concentration asymptotic behavior.
Notably, these predictions are obtained without assuming specific values of the coefficients in the mean-field adsorption model.
We therefore expect that the design rules that we have derived for programmable surfactants hold beyond the lattice models that we have simulated in this work.

\section{Discussion}

In this work, we consider a simplified, coarse-grained model of a ``programmable surfactant'' in a multicomponent biomolecular mixture.
Our central results are a set of design rules for controlling multiphase condensate morphologies, which are summarized in the wetting phase diagrams presented in \figref{fig:PD}.
Most importantly, these phase diagrams demonstrate that surprisingly low concentrations of a weakly interacting programmable surfactant can induce a transition from non-wetting (i.e., ``dispersed'') to wetting (i.e., ``docked'' or ``core--shell'') configurations.
More precisely, we find that the heterotypic interactions between the surfactant-like species and the majority component(s) of the condensed phases must exceed a relatively weak binding strength.
However, given heterotypic interactions that are slightly stronger than this threshold value, a surfactant volume fraction of only a few percent is needed to trigger the wetting transition.
These observations imply that relatively small changes to the state of the system---either small adjustments to the concentrations or heterotypic binding strengths of the surfactant-like species---can alter the equilibrium morphology of a multiphase condensate.

We find that the qualitative features of the wetting phase diagrams agree between our molecular simulation results and the predictions of two theoretical approaches.
From our molecular simulations, we predict the wetting behavior in the thermodynamic limit using measurements of the potential of mean force between condensates in a finite-size system.
In the CDFT approach, we minimize an approximate grand-potential functional to obtain predicted equilibrium concentration profiles in the vicinity of the condensate interfaces, which agree semi-quantitatively with our simulation results.
Finally, we show that a simplified adsorption model captures the key features of our detailed calculations, suggesting that the design rules that we have extracted from the wetting phase diagrams are likely to apply much more generally to related models of programmable surfactants with different molecular details.
For example, we have shown that our simulation methods and adsorption model can be applied to biomolecules with directional as opposed to isotropic interactions.

Returning to the stress granule (SG)/P-body (PB) system that motivated our model, we propose that the insights gained from our calculations can be applied directly to protein/RNA interaction networks that underlie the phase behavior of multiphase biomolecular condensates.
The key step lies in coarse-graining the interaction network to identify potential surfactant-like species, which should interact with protein/RNA components in multiple, distinct condensed phases.
Such species are likely to be situated as ``bridges'' between strongly interacting portions of the network~\cite{sanders2020competing}.
For example, in the SG/PB system~\cite{sanders2020competing}, the protein DDX6 (\figref{fig:1}A) is an obvious candidate, as it is weakly recruited to both condensates.
Our model predicts that this species should be weakly enriched at SG/PB interfaces in the endogenous system, which exhibits a stable docked morphology.
This testable prediction is also reminiscent of recent findings that certain proteins are localized to the nucleoli rim~\cite{stenstrom2020mapping}.
In future work, we will examine theoretical methods for identifying surfactant-like species on the basis of the interaction network structure and experimentally determined binding affinities and expression levels.
In light of our current results, the requirements of moderate binding strengths and low molecular concentrations suggest that this proposed mechanism of a molecular ``switch'' for controlling intracellular condensate morphologies is likely to be biologically relevant.
Further experiments are needed to test the detailed predictions of our model.

Finally, we note that this mechanism is not limited to naturally occurring biomolecular systems.
Low-concentration, surfactant-like molecular switches may also be useful for tuning multiphase morphologies in materials engineering, where approaches that do not require substantial changes to the bulk properties of the coexisting phases are similarly desirable.
For example, the phase behavior of multiphase DNA ``nanostar'' liquids~\cite{NSs} can be interpreted in terms of an interaction network, in which ``cross-linker'' nanostars can be engineered to play the role of the molecular switch.
Our model and theoretical framework can also be applied to engineer complex multiphase emulsions~\cite{sheth2020multiple}, which have broad applications including encapsulation and triggered delivery of molecular cargoes. 
These systems and similar examples of programmable soft matter~\cite{sato2020} would be ideal opportunities to test our predictions experimentally and to apply the design rules developed in this work.

\begin{acknowledgments}
  This work is supported by the National Science Foundation (DMR-2143670).
\end{acknowledgments}

\appendix
\setcounter{figure}{0}
\renewcommand{\thefigure}{A\arabic{figure}}

\section{Details of Monte Carlo simulations} 
\label{app:simulations}

In our lattice model, each molecule interacts with its six nearest neighbors according to the interaction matrix, $\bm\epsilon$.
Bivalent bridge molecules, which can engage in at most two nearest-neighbor interactions, are the sole exception to this rule.
Vacant lattice sites represent the inert solvent.
Since we are interested in investigating how the B species controls the condensate interfaces as opposed to the properties of the bulk phases, we choose to keep the homotypic N1 and N2 interactions constant in this work.
To guarantee stable $\alpha$ and $\beta$ phases, we fix these interaction strengths to be $1.5 k_{\text{B}} T$ per bond, which is stronger than the critical binding strength $\sim0.89\, \kT$ of the cubic lattice gas model~\cite{jacobs2014phase}.
The heterotypic interactions between the N1 and N2 species are set to zero to ensure immiscibility of the $\alpha$ and $\beta$ phases.
The interaction energies describing heterotypic N1--B and N2--B interactions, $\epsilon_{\text{NB}}$, and homotypic B--B interactions, $\epsilon_{\text{BB}}$, are left as free parameters.
Under these conditions, the N1 and N2 concentrations primarily affect the volume fractions of the coexisting $\alpha$, $\beta$, and dilute phases and have negligible effects on the compositions of the condensed phases.
We therefore fix the volume fractions of the N1 and N2 species to be $\phi_\text{N1}=\phi_\text{N2}=0.25$, such that the condensed phases occupy approximately half the total volume.

We implement direct-coexistence simulations using a $100 \times 8 \times 8$ lattice with periodic boundary conditions.
Simulations are carried out using the Metropolis Monte Carlo (MC) algorithm~\cite{metropolis1953}, where we attempt to exchange the positions of molecules of different types, including vacancies, at each MC move.
In each MC sweep, we attempt 6,400 moves, which is the total number of lattice sites in the simulation box. 
In simulations with bivalent bridge molecules, each B molecule has 6 orientational states.
We therefore attempt particle-swap and B-molecule rotation moves, applied to a randomly selected B molecule, with equal probability.
Each MC sweep in this case consists of twice as many moves.

The simulation-box geometry results in approximately planar interfaces between coexisting phases.
We therefore compute molecular volume-fraction profiles, $\vec\phi(z)$, as a function of the $z$ coordinate along the long dimension of the simulation box.
The interface between a condensed phase and the dilute phase is well described by a hyperbolic tangent function~\cite{HANSEN2013203}, $\phi_{\text{N1}}(z) = {(1/2) (\phi_{\text{N1}}^{(\alpha)}+\phi_{\text{N1}}^{(\text{D})})} + {(1/2) (\phi_{\text{N1}}^{(\alpha)}-\phi_{\text{N1}}^{(\text{D})}) \text{tanh}[(z-z_0) / \xi)]}$, where $\phi_{\text{N1}}^{(\alpha)}$ and $\phi_{\text{N1}}^{(\text{D})}$ are the volume fractions occupied by the molecular species N1 in the bulk $\alpha$ and dilute phases, respectively (see \figref{fig:2}A).
This expression is used to define the interfacial width, which is equal to $2\xi$.
We also define the width of the $\alpha$ condensed phase, $l_0$, based on the distance between the left and right Gibbs dividing surfaces, where $\phi_{\text{N1}}(z) = (\phi_{\text{N1}}^{(\alpha)}+\phi_{\text{N1}}^{(\text{D})})/2$.

We record volume fraction profiles and the COM distances between condensates at the same time in simulations.
We first equilibrate the system for 5000 MC sweeps, which we determine to be much longer than the equilibration time based on the COM distance fluctuations.
We then perform a production run as described in the main text.
The reference point for the PMF calculation is chosen to be $r_\text{ref} = 35\sigma$ for the isotropic bridge model and $r_{\rm ref}=33\sigma$, for the bivalent bridge model.
The target values for the COM distances during umbrella sampling are in the range $25\sigma \le r_{0,i} \le 35\sigma$ for the isotropic bridge model and $23\sigma \le r_{0,i} \le 33\sigma$ for the bivalent bridge model.
To estimate the equilibrium volume fraction profiles shown in \figref{fig:CDFT}A, we average the profiles from a simulation performed under the umbrella potential with $r_{0,i}=r_{\rm eq}$.
Importantly, this approach does not affect the equilibration of degrees of freedom orthogonal to the coordinate $r$.

\section{Finite-size scaling of the PMFs} 
\label{app:scaling}

\begin{figure}
  \includegraphics[width=8.5cm]{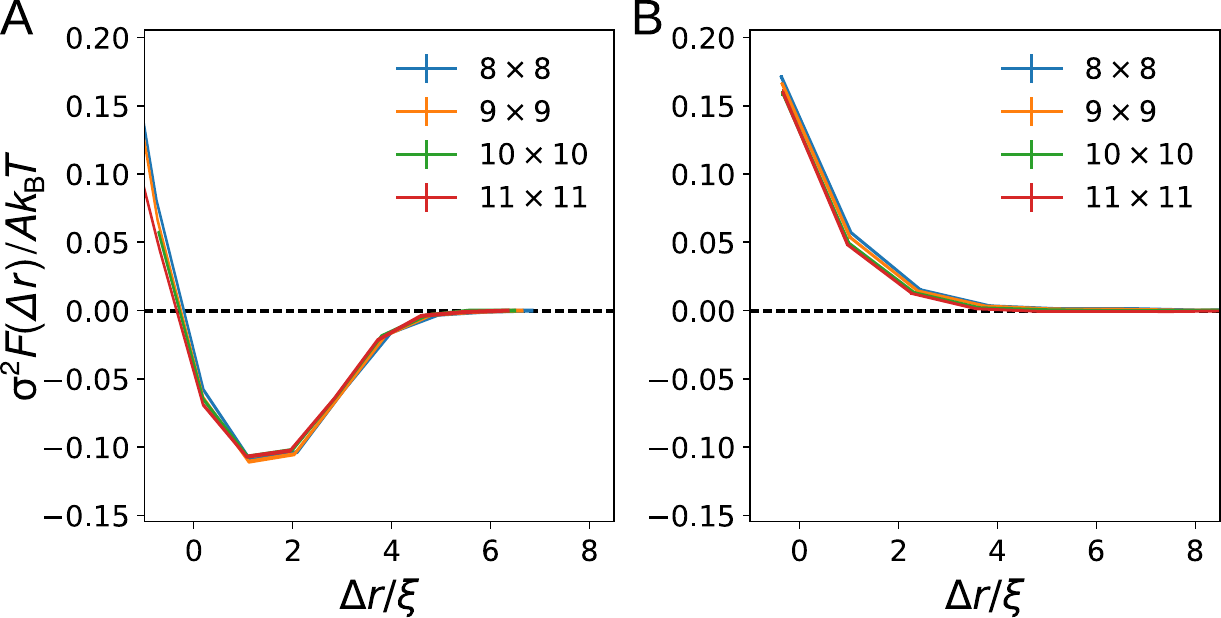}
  \caption{\textbf{Finite-size scaling of the PMFs.}  PMFs obtained for the isotropic bridge model collapse onto a single curve when scaled by the cross-sectional area, $A = \sigma^2 l^2$, in both (A)~wetting ($\epsilon_{\text{NB}} = -1k_\text{B}T$, $\epsilon_{\text{BB}} = 0$, $\phi_\text{B} = 0.1$) and (B)~non-wetting ($\phi_\text{B} = 0$) examples.  The cross-sectional dimensions of the lattice, $l \times l$, are indicated for each curve in units of $\sigma$.  Statistical errors are smaller than the line width.}
  \label{fig:scaling}
\end{figure} 

In this section, we show that the condensate morphology of the system with an arbitrary finite size, or in the thermodynamic limit, can be predicted on the basis of PMF calculations obtained from finite-size simulations.
Within a finite simulation box, the probability of distribution of the $\alpha$--$\beta$ COM distance is related to the PMF via $p(r) \propto e^{-F(r) / \kT} = e^{-A f(r) / \kT}$, where $A$ is the cross-sectional area of the simulation box and $f(r)$ is the PMF per unit area.
Consistent with this expectation, we indeed find that the PMF profiles scale with $A$ in our simulations in both wetting and non-wetting cases (\figref{fig:scaling}).
These results indicate that the PMFs capture extensive properties of the condensed-phase interfaces in our model and are not significantly influenced by the dimensions of the simulation box.
Next, we define a contact distance $r_{\text{c}} = l_0+\lambda$ beyond which we consider the condensates to be in a non-wetting configuration, such that $F(r_{\text{c}}) \simeq 0$.
Here $l_0$ depends on the width of the condensed phase while $\lambda$ is a constant.
The probability of finding the $\alpha$ and $\beta$ condensed phases in a wetting configuration in a simulation box of length $L$ can then be written as
\begin{equation}
    p_{\text{w}} = \frac{Z_{\text{w}}}{Z_{\text{w}} + Z_{\text{nw}}},
    \label{eq:finiteP}
\end{equation}
where $Z_{\text{w}} \equiv \int^{r_{\text{c}}}_0 e^{- Af(r) / \kT} dr$ and $Z_{\text{nw}} = L/2 - r_{\text{c}}$ are the partition functions associated with the wetting and non-wetting macrostates, respectively.

We now consider changing the geometry of the simulation box while keeping the concentrations of all molecular species unchanged.
As a result, the volume associated with each condensed phase, $l_0 A$, scales with the system size, while the volume fractions and compositions of the condensed phases remain constant.
Since $Z_{\text{w}}$ depends on the cross-sectional area $A$ and $Z_{\text{nw}}$ depends on the box length $L$, the wetting probability depends on both $A$ and $L$ in a finite-size simulation box.
The $L$-dependence indicates that configurational entropy plays a role in determining $p_{\text{w}}$ in a finite-size system, implying that the probability of forming a wetting interface tends to zero if the simulation box is elongated with the cross-sectional area $A$ held constant.
However, in the thermodynamic limit, both $A$ and $L$ are taken to infinity with the ratio $A^{1/2}/L$ held constant.
The wetting probability then tends to either one or zero, depending on whether the minimum of the PMF is less than zero.
If the PMF minimum is negative, then $Z_\text{w}$ scales exponentially with $A$ while $Z_\text{nw}$ scales with $A^{1/2}$; in this case, $p_\text{w} = 1$ according to \eqref{eq:finiteP}.
By contrast, if the minimum value of $F(r)$ is non-negative, then $Z_\text{w}$ decreases with $A$, and $p_\text{w} = 0$.

These arguments are easily extended to describe the morphology of spherical multiphase condensates.
The finite-size wetting probability, \eqref{eq:finiteP}, is now determined from the partition functions $Z_{\text{w}} =  4\pi \int^{r_{\text{c}}}_0 e^{- A(r) f(r) / \kT} r^2 dr$ and $Z_{\text{nw}} = V - (4\pi/3) r_{\text{c}}^3$, where the interfacial area, $A(r)$, depends on the COM distance, $r$.
The non-wetting partition function, $Z_{\text{nw}}$, represents the free volume available to a non-wetted condensate, and $V$ is the total volume of the system.
In the thermodynamic limit, an analogous scaling argument implies that the wetting behavior again depends solely on the surface-tension difference, which is related to the PMF minimum via \eqref{eq:PMFwell}.
When $-\gamma_{\alpha \beta} + 2 \gamma_{\alpha \text{D}} > 0$, the docked condensates take the shape of spherical caps~\cite{deGennes_wetting}, forming a circular interface with contact angle $\theta = \arccos (-\gamma_{\alpha \beta} / 2 \gamma_{\alpha \text{D}})$; otherwise, spherical $\alpha$ and $\beta$ condensates do not form a stable shared interface in the thermodynamic limit.

\section{CDFT phase-coexistence calculations} 
\label{app:coexistence}

We solve for phase coexistence in the regular solution model (see \secref{sec:cdft}) following the numerical strategy described in Ref.~\cite{jacobs2023theory}.
Specifically, we solve for the molecular volume fractions, $\{\vec{\phi}^{(k)}\}$, and the mole fractions of the coexisting phases, $\{x^{(k)}\}$, given the total molecular volume fractions, $\vec{\phi}_{\tmop{tot}}$.
Mass conservation requires that
\begin{equation}
  \vec{\phi}_{\text{tot}} = \sum^m_{k = 1} x^{(k)} \vec{\phi}^{(k)},
\label{eq:mass-conserv}
\end{equation}
if there are $m$ phases in coexistence.

\begin{figure*}[t]
  \includegraphics[width=\textwidth]{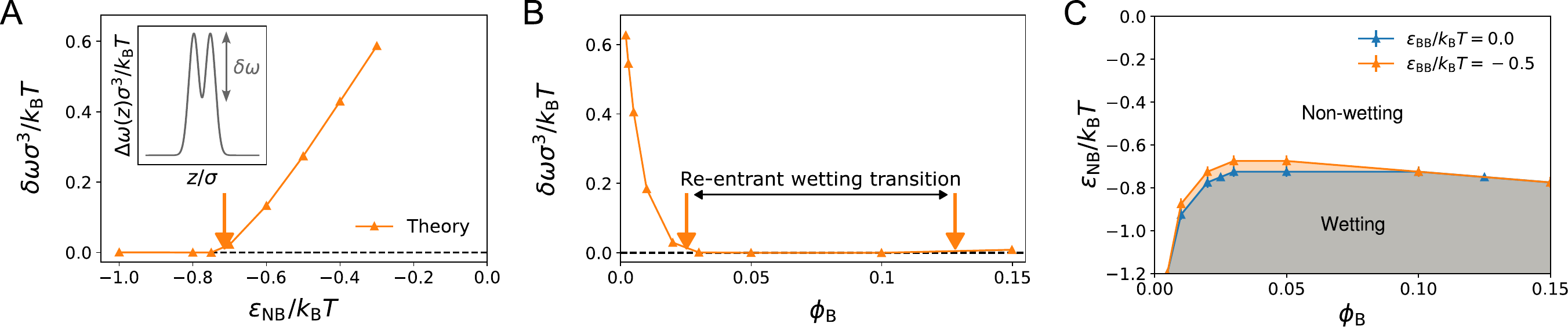}
  \caption{\textbf{Empirical criterion for locating the wetting transition in CDFT.}  (A--B)~The location of the wetting transition can be empirically estimated from CDFT by examining where the double peaks of the excess free-energy profile merge (inset).  In \textbf{B}, this method predicts a re-entrant wetting transition at constant $\epsilon_{\text{NB}} = -0.75\,\kT$.
  The calculations in \textbf{A}--\textbf{B} use the isotropic bridge model with $\epsilon_{\text{BB}} = 0$. (C)~The wetting phase diagrams predicted by this empirical criterion.}
  \label{fig:fep}
\end{figure*} 

We now consider the conditions for phase equilibrium.
The grand-potential density is related to the free-energy density via ${\omega_0 = f_0 - \sum_{i=1}^N (\partial f_0 / \partial \phi_i) \phi_i}$, where the chemical potentials of the non-solvent molecular species are ${\mu_i = \sigma^3 \partial f_0 / \partial \phi_i}$.
Coexisting phases are located at minima of $\omega_0$, ensuring that all components have equal chemical potentials in all phases.
Phase equilibrium also requires equal pressures across all coexisting phases, implying that the $\omega_0(\vec\phi)$ has the same value at all local minima.
Together, these conditions require
\begin{equation}
  \omega_0(\vec{\phi}^{(k)}) = \min \omega_0(\vec{\phi}; \vec{\mu}).
  \label{eq:phase-equilibrium}
\end{equation}
We solve \eqref{eq:phase-equilibrium} numerically by minimizing the norm of the residual errors of \eqref{eq:mass-conserv} and \eqref{eq:phase-equilibrium} iteratively.
At each iteration, we first locate the local minima of the grand potential, $\vec{\phi}^{(k)} = \arg \min_{\vec{\phi}} \omega_0(\vec{\phi};\vec{\mu})$ for all phases $k=1,\ldots,m$, using the Nelder-Mead method~\cite{nelder-mead}.
We then update $\vec{\mu}$ and $\{x^{(k)}\}$ using the modified Powell method~\cite{hybrmethod}.

Success of this optimization procedure requires that the initial estimates of $\{\vec\phi^{(k)}\}$ are not too far from the values at coexistence.
We obtain an initial guess for $\{\vec\phi^{(k)}\}$ from the convex hull method~\cite{mao2019phase,jacobs2023theory}, in which we locate the convex hull of points on a discretized $(N+1)$-dimensional free energy surface.
We initialize our optimization procedure with $N + 1$ vectors $\{\vec{\phi}^{(k)}\}$ that correspond to the vertices of the convex hull facet that encompasses $\vec\phi_{\text{tot}}$.
From the linear equation that defines this facet, we also obtain initial guesses for $\vec{\mu}$ and $\{x^{(k)}\}$.
When the number of coexisting phases $m$ is less than $N+1$, some of the $N+1$ vectors $\{\vec{\phi}^{(k)}\}$ are identical to within numerical tolerance after optimization; in this case, we restart the optimization procedure with a unique set of vectors and the corresponding $\vec{\mu}$ and $\{x^{(k)}\}$.
In this way, we determine the number of coexisting phases, $m$, as well as the molecular volume-fraction vectors, $\{\vec\phi^{(k)}\}$, and chemical potentials, $\vec\mu$, at phase coexistence.

\section{Numerical solution of the CDFT Euler--Lagrange equation}
\label{app:euler-lagrange}

To minimize the grand-potential functional in \eqref{eq:grandpotential}, we employ a numerical approach based on gradient descent.
We first discretize the $z$ coordinate, oriented perpendicular to the planar interface between phases, as $\{z_k\}$ for $k = 1, 2, \ldots, n$.
As a result, the integration in \eqref{eq:grandpotential} becomes a summation, and the grand-potential functional becomes a function of $n$ $N$-dimensional vectors $\vec{\phi}_k \equiv \vec{\phi} (z_k)$.
Using the central difference formula for differentiation, \eqref{eq:grandpotential} becomes
\begin{equation}
  \frac{\Omega}{A h} = \sum^{n - 1}_{k = 2} \! \left[ \omega_0
  (\vec{\phi}_k) \!+\! \frac{ (\vec{\phi}_{k + 1} \!-\! \vec{\phi}_{k - 1}\!)^\top \cdot \bm{m} \cdot (\vec{\phi}_{k + 1} \!-\! \vec{\phi}_{k - 1}\!)}{2 (2 h)^2} \right]\!,
  \label{eq:discretized}
\end{equation}
where $h \equiv z_{k + 1} - z_k$ is the discretization interval.
We fix the vectors $\vec\phi_k$ at the two points closest to each boundary, $k = 1, 2, n-1,$ and $n$, to be equal to the bulk phase densities.
In all calculations, we set $h=0.02\sigma$ and $n=1004$.
These choices separate the bulk-phase boundary conditions by a distance of $20\sigma$, which is much greater than the typical interfacial width (see \figref{fig:2}A).

We then apply gradient descent to minimize the discretized grand potential, \eqref{eq:discretized}, by calculating the partial derivatives $\partial \Omega / \partial \vec{\phi}_k$.
At each optimization step $l$, the densities are updated according to
\begin{equation}
  \vec{\phi}^{\;(l + 1)}_k =\vec{\phi}^{\;(l)}_k - \lambda \left(
  \frac{\partial \Omega}{\partial \vec{\phi}_k} \right)_{\!\vec\phi^{\,(l)}}\!,
\label{eq:GD}
\end{equation}
where $\lambda$ controls the step size.
We choose $10^{-3}$ as the initial value of $\lambda$ and reduce it by half if an attempted step increases the grand potential.
We initialize this optimization algorithm using an interface with a width of $4\sigma$ and a piecewise-linear spatial variation of the molecular volume fractions.
The algorithm terminates when the norm of the gradient falls below a threshold value, $10^{-3}$, at which point we take $\vec{\phi}_k$ to be the equilibrium profile.

To verify the convergence of this algorithm, we perturb the concentration profiles by moving the $\alpha$ and $\beta$ interfaces apart by $0.04\sigma$ and then restarting the optimization algorithm.
In a wetting scenario, the profile converges back to the result of the original optimization.
However, in a non-wetting scenario, the profile remains close to the perturbed profile, consistent with an unstable interface.
In practice, we compare the norms of the distances between the re-optimized, perturbed, and originally optimized profiles to verify the optimization result.

In predicted wetting cases where the CDFT surface tension is positive yet close to zero, we observe that the equilibrium excess free-energy profile becomes doubly peaked (inset of \figref{fig:fep}A).
This feature can serve as an empirical, yet practically useful, criterion for estimating the conditions for the wetting transition, since it is numerically much easier to detect this feature than to converge the calculations to the precision required to compute the surface tension difference (\figref{fig:fep}A--B).
We empirically find that the transition from singly to doubly peaked excess free-energy profiles results in reasonable agreement with the wetting phase diagram computed from our simulation results (\figref{fig:fep}C).

\section{CDFT linear path and minimum free-energy path approximations}
\label{app:CH}

In some cases, the CDFT results that we obtain by solving \eqref{eq:euler-lagrange} numerically differ qualitatively from the predictions of a ``linear path approximation'' that has appeared in previous studies of multicomponent fluids~\cite{mao2019phase,mao2020designing,sanders2020competing}.
To demonstrate the important differences between these approaches, we follow the Cahn--Hilliard approach~\cite{cahn1958free} and integrate the Euler--Lagrange equation, \eqref{eq:euler-lagrange}, to obtain
\begin{equation}
  \omega_0 [\vec\phi_{\text{eq}}(z)] - \omega_0^{(\text{D})} = \frac{1}{2}
  \sum_{i, j} \phi'_{\text{eq},i} (z) m_{i j} \phi'_{\text{eq},j} (z),
  \label{eq:integratePDE}
\end{equation}
which relates the bulk and square-gradient contributions to the excess free energy, \eqref{eq:excess-free-energy}, of the equilibrium interface.
However, to make further progress using \eqref{eq:integratePDE}, we need to know the path through concentration space that corresponds to the equilibrium interface between the coexisting phases $\alpha$ and $\beta$.
In general, this path can be described parametrically by $\vec{\phi}(\eta)$, with $0 \le \eta(z) \le 1$, $\lim_{z\rightarrow-\infty}\eta(z) = 0$, and $\lim_{z\rightarrow\infty}\eta(z) = 1$. 

Unlike the case of a binary mixture, the equilibrium path through concentration space is typically not known \textit{a priori} in a multicomponent system.
Assuming a linear path, $\vec{\phi}(\eta)=(1-\eta)\vec{\phi}^{(\alpha)}+\eta \vec{\phi}^{(\beta)}$, leads to the expression utilized in Refs.~\cite{mao2019phase,mao2020designing,sanders2020competing},
\begin{equation}
    \gamma_{\alpha\beta} = \sqrt{2(\phi^{(\beta)}_i\!-\!\phi^{(\alpha)}_i)m_{ij}(\phi^{(\beta)}_j\!-\!\phi^{(\alpha)}_j)}\!\int^1_0 [\Delta\omega_0(\eta)]^{1/2} d\eta,
    \label{eq:CH}
\end{equation} 
where $\Delta\omega_0(\eta) \equiv \omega_0(\vec{\phi}(\eta)) - \omega_0^{(\text{D})}$.
However, this linear-path assumption implies that the concentration of the B molecule cannot be greater at the interface than it is in the coexisting bulk phases, which is generally inconsistent with our simulation results.
For this reason, \eqref{eq:CH} fails to predict wetting configurations for all conditions that we consider in this work.

A plausible alternative approximation is to find the path through concentration space that minimizes the bulk contribution to the excess free energy.
This approximation leads to a minimum-free-energy path (MFEP) assumption for $\vec{\phi}(\eta)$.
More precisely, the MFEP minimizes the integral $\int_0^1 \Delta\omega_0(\vec{\phi}(\eta))\,d\eta$ and thus passes through a saddle point on the grand-potential landscape between the bulk-phase concentrations $\vec\phi^{(\alpha)}$ and $\vec\phi^{(\beta)}$.
In practice, we calculate the MFEP using a direct implementation of the zero-temperature string method~\cite{string2007}.
(For a discussion of the algorithmic details of this method, we refer the reader to Ref.~\cite{string2007}.)
Following the terminology in Ref.~\cite{string2007}, we use a total of $N=100$ points on a string between two local minima on the bulk excess free energy surface.
In the ``evolution step'' of the algorithm, the points evolve according to the gradient descent method, as in \eqref{eq:GD}, where $\Delta t$ controls the step size.
Here we use the forward Euler method with $\Delta t=5\times10^{-4}$.
Then in the ``reparameterization step'', we reparameterize the string such that the $N$ points are equally spaced with respect to arc length along the string.
We check for convergence by measuring the norm of the displacement of all points from their positions in the previous iteration, and we use a convergence tolerance of $\text{TOL}=10^{-4}$.

By contrast with the linear-path assumption, the MFEP tends to predict enrichment of the B component at the interface whenever $\epsilon_{\text{NB}} < 0$.
Nonetheless, the MFEP approximation can misclassify the interface as non-wetting in many cases, particularly when the path is predicted to pass through the dilute phase (see \figref{fig:CDFT}D--E).
The numerically determined equilibrium path, $\vec{\phi}_{\text{eq}}(\eta)$, is intermediate between the linear path and the MFEP.
We therefore interpret the equilibrium path as resulting from a competition between the bulk and square-gradient contributions to the grand-potential functional in our multicomponent model.
For this reason, the optimal path is sensitive to the $\bm{m}$-matrix, making our regularization approach (see \secref{sec:cdft}) an essential and nontrivial aspect of our CDFT calculations.

\section{Fluctuating-interface adsorption model}
\label{app:adsorption}

In this section, we detail the essential steps to bridge the gaps between \eqref{eq:difference} and the subsequent results presented in \secref{sec:adsorption} of the main text.
Explicitly expressing the wetting condition in \eqref{eq:difference} leads to
\begin{equation}
  (\phi^{(\text{D})}_\text{B})^2e^{2a_{\rm \alpha D}}+\phi^{(\text{D})}_\text{B} (2e^{a_{\rm \alpha D}} \!-\! e^{a_{\rm \alpha \beta}-\Delta s}) + 1 \!-\! e^{-\Delta s} < 0.
  \label{eq:quadratic}
\end{equation}
\eqref{eq:quadratic} only has solutions for $\phi^{(\text{D})}_\text{B}$ when the discriminant of this quadratic function is positive, such that $(2e^{a_{\rm \alpha D}}-e^{a_{\rm \alpha\beta}-\Delta s})^2-4e^{2a_{\rm \alpha D}}(1-e^{-\Delta s})>0$.
Setting the quadratic function to 0 and considering $e^{a_{\alpha \beta}-a_{\alpha \text{D}}}\gg e^{\Delta s}$, we obtain the asymptotic formula for the low-concentration phase boundary, \eqref{eq:boundary}.

Both $a_{\rm \alpha D}$ and $a_{\alpha \beta}$ are expected to be proportional to the N--B binding energy, $\epsilon_\text{NB}$.
In addition, physical values of $\phi^{(\text{D})}_\text{B}$ must be between 0 and 1.
With these conditions, \eqref{eq:quadratic} allows us to predict the weakest N--B binding strength for a wetting interface.
Setting the discriminant to 0, the volume fraction at this binding strength is
\begin{equation}
  \phi^{(\text{D})}_\text{B}\big|^*=e^{-a_{\rm \alpha D}}\sqrt{1-e^{-\Delta s}}.
  \label{eq:phi_ltbs}
\end{equation}
At the weakest binding strength, we have $a_{\rm \alpha D}^* - a_{\rm \alpha \beta}^*=\ln (1-\sqrt{1-e^{-\Delta s}})/2$, from which we obtain \eqref{eq:ltbs}.
Then, by making the approximation ${a_{\rm \alpha \beta} \simeq 2 a_{\rm \alpha D}}$, we are able to express these quantities in terms of $\Delta s$ at the weakest N--B binding strength, $|\epsilon_{\text{NB}}^*|$, on the wetting phase boundary,
\begin{align}
    a^*_{\rm \alpha D} &= \ln 2+\ln [e^{\Delta s}+\sqrt{e^{\Delta s}(e^{\Delta s}-1)}],
    \label{eq:ltbs_2} \\
    K^*_{\rm \alpha D} &= \sqrt{1-e^{-\Delta s}},
    \label{eq:K_star} \\
    \phi^{({\rm i, \alpha \beta})}_\text{B} \big|^* &= 2\phi^{({\rm i, \alpha D})}_\text{B} \big|^*.
    \label{eq:twice}
\end{align}
According to \eqref{eq:ltbs_2}, the minimum binding strength $|\epsilon_{\text{NB}}^*|$ increases with $\Delta s$, since $a_{\rm \alpha D}\propto|\epsilon_{\text{NB}}|$.
As $\Delta s$ increases, the B-species volume fraction in the dilute phase, $\phi^{(\text{D})}_\text{B}\big|^*$, first increases until reaching a maximum value of 0.125 at $K^*_{\rm \alpha D}=0.5$, as evidenced by substitution of \eqref{eq:K_star} into \eqref{eq:phi_ltbs_2}.
$\phi^{(\text{D})}_\text{B}\big|^*$ then decreases with $\Delta s$ at larger $\Delta s$.
The wetting phase boundary is always non-monotonic for a positive $\Delta s$.
These observations suggest that an accurate treatment of the capillary fluctuations, quantified here by $\Delta s$, is essential for obtaining an accurate prediction of $|\epsilon_{\text{NB}}^*|$; this interpretation is consistent with the quantitative differences between our CDFT and simulation results in \figref{fig:PD}A-B.
We can also estimate $\phi^{(\text{D})}_\text{B}|^*$ by computing $\Delta s$ directly from our simulated PMF at zero B-species concentration, where $k_\text{B} T \Delta s / \sigma^2 = -(-\gamma_{\rm \alpha \beta}+2\gamma_{\rm \alpha D}) \simeq F(\Delta r=0) / A$.
From \figref{fig:scaling}B, we find $\Delta s \approx 0.15$, which, according to \eqref{eq:phi_ltbs_2}, suggests that $\phi^{(\text{D})}_\text{B}|^* \simeq 0.12$.
This prediction is reasonable given our simulation results (see \figref{fig:lm}B).

We can similarly apply this framework to describe the bivalent bridge model.
To account for the entropic penalty of aligning B molecules at the interface, we add a density-dependent term to $\Delta S$ in \eqref{eq:lm_excess}, $\Delta S(\phi^{(\text{i})}_{\text{B}}) / A k_\text{B} = -\phi^{(\text{i})}_{\text{B}} \ln \phi^{(\text{i})}_{\text{B}} - (1-\phi^{(\text{i})}_{\text{B}}) \ln (1-\phi^{(\text{i})}_{\text{B}}) + s + (-\ln{6})\phi^{(\text{i})}_{\text{B}}$, assuming that the B molecules at the interface are all aligned in the correct orientation with their binding sites pointed into the condensed phases.
This modification is equivalent to decreasing $a$ by $\ln 6$.
As a result, $\phi^{(\text{D})}_\text{B}\big|^*$ increases by a factor of 6 according to \eqref{eq:phi_ltbs}, explaining why we do not observe the re-entrant phase behavior in our simulations at bridge volume fractions up to $\phi_{\rm B}=0.3$.

To extend this adsorption model to incorporate homotypic B--B interactions, we modify the mean-field expressions for the enthalpic contribution to the surface excess grand potential and the chemical potential in \eqref{eq:lm_excess},
\begin{align}
  h(\phi^{(\rm i)}_\text{B}) &= -a'(\phi^{(\rm i)}_\text{B})^2-a \phi^{(\rm i)}_\text{B} +b,
  \label{eq:c_1} \\
  \mu_\text{B}/k_\text{B} T &= \ln \phi^{(\text{D})}_\text{B}-c\phi^{(\text{D})}_\text{B},
  \label{eq:c_2}
\end{align}
where $a'$ and $c$ are positive constants that are expected to be proportional to $|\epsilon_\text{BB}|$.
We can then show that these additional terms in \eqref{eq:c_1} and \eqref{eq:c_2} lead to an increase in $(-\gamma_{\alpha\beta}+2\gamma_{\rm \alpha D})|^*$ by considering perturbations to the $\epsilon_\text{BB}=0$ case.
Comparison with \eqref{eq:lm_excess} shows that these additional terms effectively alter the parameter $a$ in the original model by
\begin{equation}
  \Delta a = a'\phi^{(\rm i)}_\text{B}-c\phi^{(\rm D)}_\text{B}.
\end{equation}
According to \eqref{eq:gamma}, introducing the perturbation $\Delta a$ changes the surface tension by an amount
\begin{equation}
  \Delta \gamma = -\phi^{(\rm i)}_\text{B}\Delta a = -\phi^{(\rm i)}_\text{B}(a'\phi^{(\rm i)}_\text{B}-c\phi^{(\rm D)}_\text{B}),
\end{equation}
and thus the surface tension difference by an amount
\begin{equation}
    \begin{aligned}
        \Delta (-\gamma_{\rm \alpha \beta}+2\gamma_{\rm \alpha D})=&[(\phi^{({\rm i, \alpha \beta})}_\text{B})^2-2(\phi^{({\rm i, \alpha D})}_\text{B})^2]a'\\
        &-(\phi^{({\rm i, \alpha \beta})}_\text{B}-2\phi^{({\rm i, \alpha D})}_\text{B})\phi^{(\text{D})}_\text{B}c.
    \end{aligned}
\label{eq:delta_gamma}
\end{equation}
Again assuming that ${a_{\rm \alpha \beta} \simeq 2 a_{\rm \alpha D}}$, we apply these results to the minimum binding strength, $|\epsilon_{\text{NB}}^*|$, on the $\epsilon_{\text{BB}} = 0$ wetting phase boundary.
Simplifying \eqref{eq:delta_gamma} using \eqref{eq:iso} and \eqref{eq:twice}, we find that the term involving the parameter $c$ cancels out, and we arrive at \eqref{eq:delta_gamma*}.
The fact that $\Delta (-\gamma_{\rm \alpha \beta}+2\gamma_{\rm \alpha D})|^*$ is positive, regardless of specific choices for the parameters $a'$ and $c$, indicates that the minimum binding strength $|\epsilon^*_\text{NB}|$ is reduced compared to the $\epsilon_\text{BB}=0$ scenario.
Thus, this model predicts that $\epsilon_{\text{NB}}^*$ increases for $\epsilon_{\text{BB}} < 0$, consistent with our simulation results.

\end{document}